\documentclass[a4paper,notitlepage,onecolumn]{article}

\usepackage{graphicx}
\usepackage{psfrag}
\usepackage{amsmath}
\usepackage{dsfont}
\usepackage{amsmath}
\usepackage{amssymb}
\usepackage{amsthm}
\usepackage{subfig}
\usepackage[left=3cm,top=2.5cm,right=3cm,bottom=3.3cm]{geometry}

\def\1{\mathbf{1}}
\def\0{\mathbf{0}}

\def\Z{\mathds{Z}}
\def\N{\mathds{N}}
\def\R{\mathds{R}}
\def\C{\mathds{C}}
\def\P{\mathds{P}}

\def\I{{\cal I}}
\def\K{{\cal K}}

\def\a{{\bf a}}
\def\b{{\bf b}}
\def\x{{\bf x}}
\def\u{{\bf u}}
\def\k{{\bf k}}
\def\h{{\bf h}}
\def\e{{\bf e}}
\def\f{{\bf f}}
\def\b{{\bf b}}
\def\v{{\bf v}}

\def\t{{\bf t}}
\def\y{{\bf y}}
\def\K{{\bf K}}
\def\H{{\bf H}}
\def\I{{\bf I}}
\def\P{{\bf P}}

\def\A{{\bf A}}
\def\B{{\bf B}}
\def\Rb{{\bf R}}
\def\S{{\bf S}}
\def\J{{\bf J}}

\def\z{{\bf z}}
\def\Rb{{\bf R}}

\def\phib{\boldsymbol{\phi}}
\def\psib{\boldsymbol{\psi}}
\def\varphib{\boldsymbol{\varphi}}
\def\Gammab{\boldsymbol{\Gamma}}

\def\xib{\boldsymbol{\xi}}

\def\trace{{\rm tr}\,}
\def\E{\,\mathds{E}\,}

\def\MSE{\text{MSE}}

\newtheorem{theorem}{Theorem}[section]
\newtheorem{corollary}[theorem]{Corollary}
\newtheorem{lemma}[theorem]{Lemma}
\newtheorem{proposition}[theorem]{Proposition}

\newtheorem{remark}[theorem]{Remark}
\newtheorem{example}[theorem]{Example}
\numberwithin{equation}{section}

\graphicspath{{figures/}}

\begin{document}

\title{Distributed Estimation over Wireless Sensor Networks\\ with Packet Losses
\thanks{C.~Fischione and A.~Sangiovanni-Vincentelli wish to acknowledge the
support of the NSF ITR CHESS and the GSRC. The work by A.~Speranzon
was partially supported by the European Commission through the Marie
Curie Transfer of Knowledge project BRIDGET (MKTD-CD 2005 029961).
C.~Fischione and A.~Speranzon acknowledge the support of the San
Francisco Italian Institute of Culture by the Science \&
Technology Attach\'e T.~Scapolla. The work by K. H. Johansson was partially funded by the Swedish Foundation for Strategic Research and by the Swedish Research Council.}}
\author{C.~Fischione\thanks{C.~Fischione and A.~Sangiovanni-Vincentelli are with University of California at Berkeley, CA. E-mail:
\texttt{\{fischion,alberto\}@eecs.berkeley.edu}} \and A.~Speranzon\thanks{A.~Speranzon is
with Unilever R\&D Port Sunlight, CH63 3JW, United Kingdom. E- mail:
\texttt{alberto.speranzon@gmail.com}.} \and K.~H.~Johansson\thanks{K.~H.~Johansson is with the
ACCESS Linnaeus Center, Electrical Engineering, Royal Institute of Technology, 100-44 Stockholm, Sweden. E-mail: \texttt{kallej@ee.kth.se}}\and A. Sangiovanni-Vincentelli$^\dag$}

\maketitle
\begin{abstract}
A distributed adaptive algorithm to estimate a time-varying
signal, measured by a wireless sensor network, is designed and
analyzed. One of the major features of the algorithm is that no
central coordination among the nodes needs to be assumed. The
measurements taken by the nodes of the network are affected by
noise, and the communication among the nodes is subject to packet
losses. Nodes exchange local estimates and measurements with
neighboring nodes. Each node of the network locally computes
adaptive weights that minimize the estimation error variance.
Decentralized conditions on the weights, needed for the
convergence of the estimation error throughout the overall
network, are presented. A Lipschitz optimization problem is posed
to guarantee stability and the minimization of the variance. An
efficient strategy to distribute the computation of the optimal
solution is investigated. A theoretical performance analysis of
the distributed algorithm is carried out both in the presence of
perfect and lossy links. Numerical simulations illustrate
performance for various network topologies and packet loss
probabilities.
\end{abstract}

\begin{center}
\end{center}


\section{Introduction}
\label{sec:introduction}

Monitoring physical variables is a typical task performed by
wireless sensor networks (WSNs). Accurate estimation of these
variables is a major need for many applications, spanning from
traffic control, industrial manufacturing automation, environment
monitoring, to security
systems~\cite{EGH99}--\nocite{GK03}\cite{SIDSP06}. However, nodes
of WSNs have limitations, such as scarcity of energy supply, lightweight
processing and communication functionalities, with the
consequence that sensed data are affected by bias and noise, and
transmission is subject to interference, which results in
corrupted data (packet loss). Estimation algorithms must be
designed to cope with these adverse conditions, while offering
high accuracy.

There are two main estimation strategies for WSNs. A traditional
approach consists in letting nodes sense the environment and then
report data to a central unit, which extracts the desired physical
variable and sends the estimate to each local node for local
action. However, this approach has strong limitations: large
amount of communication resources (radio power, bandwidth,
routing, etc.) have to be managed for the transmission of
information from nodes to the central unit and vice versa, which
reduces the nodes' lifetime. An alternative approach, which we
investigate in this paper, enables each node to locally produce
accurate estimates taking advantage of data exchanged with only
neighboring nodes. Indeed, wireless communication makes it natural
to exploit cooperative strategies, as it has been already used for
coding and transmission~\cite{SIDSP06,JSAC06}. The challenge of
distributed estimation is that local processing must be carefully
designed to avoid heavy computations and spreading of local errors
throughout the network.

In this paper we consider the design and analysis of a distributed
estimation algorithm. Specifically, a time-varying signal is
jointly tracked by the nodes of a WSN, in which each node computes
an estimate as a weighted sum of its own and its neighbors'
measurements and estimates. The distributed estimator features
three particular characteristics: it is robust to packet losses,
it does not rely on a model of the signal to track, and it uses
filter coefficients that adapt to the changing network topology
caused by packet losses. We show that the estimation problem has a
distributed implementation. It is argued that the estimator
exhibits high accuracy, in term of estimation error variance, even
in the presence of severe packet losses, if the signal to track is
varying slowly.

\subsection{Related Work}

The estimator presented in this paper is related to recent
contributions on low-pass filtering by diffusion mechanisms,
e.g.,~\cite{paperCDC06}--\nocite{paperICC07,Morse03-1,OM04,ConsensusAlberto,XiaoBoydLall,SOSM-CDC05}\cite{Olfati05},
where each node of the network obtains the average of the initial
samples collected by nodes. In~\cite{XiaoBoydKim,XiaoBoyd} the
authors study a distributed average computation of a time-varying
signal, when the signal is affected by a zero-mean noise.
Distributed filtering using model-based approaches is studied in
various wireless network contexts,
e.g.,~\cite{SOSM-CDC05},~\cite{shang}--\nocite{GK-CDC06,OlfatiCDC07}\cite{Alriksson+07}.
In particular, distributed Kalman filters and more recently a
combination of the diffusion mechanism with distributed Kalman
filtering have been proposed, e.g.,~\cite{CarliChiuso+}.

In~\cite{JSAC07}, we have presented a distributed estimator to
track a time-varying signal without relaying on a model of the
signal to track, in contrast to model-based approaches,
e.g.,~\cite{SOSM-CDC05, GK-CDC06}. The approach is novel,
since~\cite{OM04}--\nocite{ConsensusAlberto}\cite{XiaoBoydLall}
are limited to averaging initial samples. Compared
to~\cite{XiaoBoydLall}--\cite{XiaoBoyd}
and~\cite{OlfatiCDC07}--\cite{CarliChiuso+}, we do not rely on the
Laplacian matrix associated to the communication graph. Our filter
parameters are computed through distributed algorithms which adapt
to the network topology and packet losses, whereas for
example~\cite{XiaoBoydKim} and~\cite{XiaoBoyd} rely on centralized
algorithms for designing the filters. The distributed estimator
proposed in this paper features better estimates when compared to
similar distributed algorithms presented in the literature, but at
the cost of a slightly increased computational complexity. With
respect to our earlier work~\cite{JSAC07}, here we provide a major
extension because we take into account lossy wireless
communications. Packet losses require a substantial redesign and
performance characterization of the filter proposed
in~\cite{JSAC07}. In this paper, we explicitly consider the effect
of packet losses (both i.i.d. and non-identical) in the design of
the adaptive weights so that the estimation error is guaranteed to
converge for any packet loss realization. The distributed minimum
variance estimator uses a Lipschitz optimization problem to
distribute the centralized stability constrains, whose
characterization is completely new. We devise a new algorithm to
distribute efficiently the computation of the solution of the
Lipschitz problem. An original analysis of the bounds on the
estimation error variance as function of the packet loss
probability and number of nodes is also discussed. We introduce
some examples to show how such bounds can be refined significantly
when the network topology is modelled by a graph in the class of
finite Cayley graphs.

The remainder of the paper is organized as follows: In
Section~\ref{sec:problem_formulation} we review and extend the
problem posed in~\cite{JSAC07} considering packet losses. In
Section~\ref{sec:distrib_min_var_estim} we deal with the design of
the optimal adaptive weights that minimize the estimation error
variance. In Section~\ref{sec:computation_of_psi} we show how to
compute efficiently some thresholds needed to bound the norm of
the estimation error. In Section~\ref{sec:performance_analysis} we
determine bounds on the estimation error variance achieved by the
proposed algorithm. Monte Carlo simulations are reported in
Section~\ref{sec:simulations} to illustrate the performance of the
proposed algorithm. Conclusions are drawn in
Section~\ref{sec:conclusions}.

%
%
%
%
%

\subsection{Notation}
\label{subsection:notation} Given a stochastic variable $x$, $\E
x$ denotes its expected value. With $\E_y x$ we mean that the
expected value is taken with respect to the probability density
function of $y$.  We keep
explicit the time dependence to remind the reader that the
realization is given at time $t$. With $\|\cdot\|$ we denote the
$\ell^2$-norm of a vector or the spectral norm of a matrix. Given
a matrix $\A$, $\ell_{m}(\A)$ and $\ell_{M}(\A)$ denote the
minimum and maximum eigenvalue (with respect to the absolute value
of their real part), respectively, and its largest singular value
is denoted by ${\gamma}(\A)$. Given the matrix $\B$, $\A \circ \B$
is the Hadamard (element-wise) product between $\A$ and $\B$. With
$\a \preceq \b$ and $\a \succeq \b$ denote the element-wise
inequalities. With $\I$ and $\1$ we denote the identity matrix and
the vector $(1,\dots,1)^T$, respectively, whose dimensions are
clear from the context. Let $\N_0=\N\cup\{0\}$. To keep the
notation lighter, the time dependence of the variables and
parameters is not explicitly indicated, when this does not create
misunderstandings.

\section{Problem Formulation}
\label{sec:problem_formulation}

Consider a WSN with $N>1$ sensor nodes. At every time instant,
each sensor in the network takes a noisy measure $u_i(t)$ of a
scalar signal $d(t)$, namely $u_i(t)=d(t)+v_i(t)$, for $t\in\N_0$
and for all $i=1,\dots,N$. We assume that $v_i(t)$, for all~$i$,
are normally distributed with zero mean and variance $\sigma^2$
and that $\E v_i(t) v_j(t)=0$ for all $t\in\N_0$.

We model the network as a weighted graph. In particular we consider
a graph, $\mathcal{G}(t)=(\mathcal{V},\mathcal{E})$, where
$\mathcal{V}=\{1,\dots,N\}$ is the vertex set and
$\mathcal{E}\subseteq \mathcal{V}\times\mathcal{V}$ is the edge set.

The set of neighbors of node $i\in\mathcal{V}$ plus node $i$ is
denoted as
$$
    \mathcal{N}_i=\{j \in \mathcal{V} :
    (j,i)\in\mathcal{E}\}\cup\{i\}\,,
$$
Namely $\mathcal{N}_i$ is the set containing the maximum number of
neighbors a node~$i$ can have, including itself.

Every node broadcasts data packets, so that these packets can be
received by any other node in the communication range. Packets may
be dropped because of bad channel conditions or radio
interference. Let $\phi_{ij}(t)$, with $i\neq j$, be a binary
random variable associated to the packet losses from node $i$ to
$j$ at time $t$~\cite{Stuber}. This random variable is defined on
the probability space $(\Omega,\mathcal{F},\Pr)$, where
$\Omega=\{0,1\}$, $\mathcal{F}$ is a $\sigma$-algebra of subsets
of $\Omega$ and $\Pr(.)$ a probability measure. For $i\neq j$, we
assume that the random variables $\phi_{ij}(t)$ are independent
with probability mass function:
\begin{align*}
    \Pr(\phi_{ij}(t)=1) &= p_{ij} \,,\\
    \Pr(\phi_{ij}(t)=0) &= q_{ij} = 1-p_{ij} \,,
\end{align*}
where $p_{ij}\in[0,1]$ denotes the successful packet reception
probability. Clearly that $p_{ii}=1$, since information locally
available is not subject to packet losses. Note also that
$p_{ji}=0$ if the packet sent from node $j$ to $i$ collided at
node $i$ due to too much wireless interferences, or if the
wireless channel of the link from $j$ to $i$ is under deep fading,
or if $i$ is too far from node $j$ to receive packets from node
$j$. The packet reception probabilities are assumed to be
independent among links, and independent from past packet losses.
These assumptions are natural when the coherence time of the
wireless channel is small if compared to the typical communication
rate of data packets over WSNs~\cite{Stuber,ieee802154}.

We assume each node~$i$ computes an estimate $x_i(t)$ of
$d(t)$ by taking a linear combination of its own and of its
neighbors' estimates and measurements. Define $\mathbf{x}_i(t) =
(x_{1}(t),\dots,x_{N}(t))^T$ and similarly $\mathbf{u}_i(t) =
(u_{1}(t),\dots,u_{N}(t))^T$, then each node computes
\begin{align}\label{eq:system_dyn}
    x_i(t)=(\k_{i}\phib_i)^T(t)\:\mathbf{x}_i(t-1)+(\h_i\phib_i)^T(t)\:\mathbf{u}_i(t)\,,
\end{align}
with $\x_i(0) = \u_i(0)$, and where
$$
    (\k_{i}\phib_i)(t) = \k_i^T(t) \circ \phib_i^T(t) =
    \left(k_{1}(t),k_{2}(t),\dots,k_{N}(t)\right)^T \circ
    \left(\phi_{1}(t),\phi_{2}(t),\dots,\phi_{N}(t)\right)^T\,.
$$
with $\k_i^T(t) \in\R^{N} \times 1$, in which the $j$-th element
is the weight coefficient used by node $i$ for information coming
from node $j$ at time $t$, and  $\phib_i(t)\in\R^{N} \times 1$
denotes the vector of the packet reception process as seen from
node $i$ with respect to all nodes of the network. Specifically,
the $j$th element of $\phib_i(t)$, with $j\neq i$, be
$\phi_{ij}(t)$. Let $\varphib_{i|t} \in\Omega^N $ denotes a
realization of the process $\phib_i(t)$ at time $t$. Notice that
at a given time instant, the $j$-th component of $\varphib_{i|t}$
is zero if no data packets are received from node $j$. Let
$\mathcal{N}_{\varphib_i}=\{j \in \mathcal{N}_i: \varphib_{i_j|t}
\neq 0\}$, namely a such set collects the nodes communicating with
node~$i$ at time $t$. The number of nodes in the set is
$|\mathcal{N}_{\varphib_i}|=\varphib_{i|t}^T\varphib_{i|t}$.

The vector $(\h_i\phib_i)(t) \in\R^{N \times 1} $ and $\h_i^T(t) \in\R^{N \times 1}$ are constructed from the elements $h_{ij}(t)$, similarly to $(\k_{i}\phib_i)(t)$.

\section{Distributed Minimum Variance Estimator}
\label{sec:distrib_min_var_estim} In this section we
describe how each node computes adaptive weights to
minimize its estimation error variance.

\subsection{Estimation Error}

Define the estimation error at node~$i$ as $e_i(t) = x_i(t) - d(t)\1$.
Introduce $\delta(t) = d(t)- d(t-1)$, then the expected error with respect to the measurement noise $\mathbf{v}_i(t)=(v_1(t),\dots,v_{N}(t))^T$ is given by
\begin{align}
    \E_\v e_i(t) = (\k_{i}\phib_i)^T(t)\: \E_v \mathbf{e}_i(t-1) +
    d(t)\left((\k_{i}\phib_i)^T(t)\:\1 + (\h_{i}\phib_i)^T(t)\:\1 - 1
    \right)- \delta(t)(\k_{i}\phib_i)^T(t) \1\,,
    \label{eq:err_dyn}
\end{align}
where we set $\mathbf{e}_i(t)=(e_{1}(t),\dots,e_{N}(t))^T$.

Typically one is interested in designing an unbiased estimator.
Notice, however that in~\eqref{eq:err_dyn} the expected error
depends on both unknowns $d(t)$ and $\delta(t)$.  The following
condition eliminates the dependence from $d(t)$:
\begin{align}
    (\k_{i}\phib_i)^T(t)\:\1 + (\h_{i}\phib_i)^T(t)\:\1 =
    1\,,
    \label{eq:cond_1}
\end{align}
for any possible realization of the packet loss process
$\phib_i(t)$. Note that~\eqref{eq:cond_1} holds both in the
presence of packet i.i.d. losses , and in the presence of
non-identical losses. The term $\delta(t)$ can be removed by
imposing that
\begin{align}
    (\k_{i}\phib_i)^T(t)\:\1 = 0\,,
    \label{eq:cond_2}
\end{align}
for any possible realization of the packet loss process
$\phib_i(t)$. By imposing constraints~\eqref{eq:cond_1}
and~\eqref{eq:cond_2}, the unknown terms would disappear from the
expected error equation, so the minimum variance estimator would
be such that $(\k_{i}\phib_i)^T(t)=\0$ and $(\h_{i}\phib_i)^T(t) =
1/|\mathcal{N}_{\varphib_i}|\1$, where
$|\mathcal{N}_{\varphib_i}|$ is the number of neighbors, including
node~$i$, that are successfully communicating with node~$i$ at
time~$t$. We will show in the next sections that by imposing only
constraint~\eqref{eq:cond_1} we are able to design an estimator
that has lower variance than one that also obey~\eqref{eq:cond_2}.
The price paid for better performance is a biased estimator.
However, assuming that $d(t)$ is slowly varying (or that the
sampling frequency is high enough with resect to the variation of
the signal), the bias is negligible and the proposed estimator
outperforms the unbiased one in terms of the estimation error
variance. This can also be understood from an intuitive point of
view: having $(\k_{i}\phib_i)^T(t)=\0$ means that nodes are
disregarding previous estimates and are just using current
measurements, which are typically corrupted by high noise. Having
a term that also weights previous estimates allows us to increment
the total available information at each node, obtaining a much
lower estimation error variance.

Notice that although~\eqref{eq:cond_1} holds, and thus $d(t)$ is
eliminated from~\eqref{eq:err_dyn}, we need further conditions on
vector $(\k_{i}\phib_i)(t)$ in order to guarantee that the
expected error asymptotically decreases.

\subsection{Convergence of Estimation Error}

In this subsection we derive conditions on the weights that ensure
that the estimation error decreasing over time, regardless the
measurement noise and the packet loss processes that affect the
system. In particular, we want to determine conditions on the
weights so that $|\E_{\phib}\E_\v e_i(t)|\rightarrow 0$ as
$t\rightarrow +\infty$.

Assume that constraint~\eqref{eq:cond_1} holds, and consider the
expected value with respect to the packet loss process
of~\eqref{eq:err_dyn}, then
$$
    \E_{\phib} \E_\v e_i(t) = \E_{\phib} (\k_i^T(t) \circ \phib_i(t)^T)
    \E_{\phib} \E_\v \mathbf{e}_i(t-1) - \delta(t) \E_{\phib} (\h_i^T(t) \circ \phib_i(t)^T)\:\1 \,,
$$
where we have used the fact that the packet losses at time $t$ are
independent from the preceding time instants, so that $\E_{\phib}(
\k_i^T(t) \circ \phib_i(t)^T \E_\v \mathbf{e}_i(t-1)) =\E_{\phib}
(\k_i^T(t) \circ \phib_i(t)^T)  \E_{\phib} \E_\v
\mathbf{e}_i(t-1)$. It is clear that the evolution of
$\E_{\phib}\E_\v e_i(t)$ depends on the overall error vector
$\mathbf{e}_i(t-1)$, namely, the error at the local node depends
on the estimation error of neighboring nodes. We thus need a set
of other $|\mathcal{N}_i|-1$ equations to describe the estimation
error of all nodes in $\mathcal{N}_i$. Obviously, each new
equation will depend on the estimation error of nodes that are two
hops from node~$i$, and so on. The full network will be considered
in this process of adding equations. Let $\mathbf{e}(t) \in
\R^{N}$ be the estimation error of the overall network, then
\begin{align}
    \E_{\phib} \E_\v \mathbf{e}(t) = \E_{\phib} (\K(t) \circ \Phi(t)) \E_\phi
    \E_v \mathbf{e}(t-1) - \delta(t) \E_{\phib} (\H(t) \circ \Phi(t)) \1 \,,
    \label{eq:error1}
\end{align}
where $\K(t) = \left( \k_1(t), \k_2(t), \dots, \k_N(t) \right)^T$ is the matrix whose rows are the vectors $\k_i^T(t)$, $i=1,\dots,N$,
and $\Phi(t) = \left( \phib_1(t), \phib_2(t), \dots, \phib_N(t) \right)^T$ is the matrix whose rows are the vectors $\phib_i^T(t)$, $i=1,\dots,N$. Let $\varphib_{|t}$ be a realization of $\Phi(t)$ at time $t$, namely $\varphib_{|t} = \left( \varphib_{1|t}, \varphib_{2|t}, \dots, \varphib_{N|t} \right)^T$. The following result holds:
\begin{proposition}
\label{prop:error-stability} Consider the
system~\eqref{eq:error1} and assume that
\begin{itemize}
\item[(i)]  $\gamma(\K(t) \circ
    \varphib_{|t})\leq\gamma_{\max}<1$ for all $t\in \N_0$ and
    for each and every packet loss realization $\varphib_{|t}$
    of $\Phi(t)$.
\item[(ii)] $|\delta(t)|<\Delta$ for all $t\in \N_0$.
\end{itemize}
Then, considering independent packet losses, we have that
\begin{align}
    \lim_{t\rightarrow +\infty}  \|\E_{\phib} \E_\v \mathbf{e}(t)\|
    \leq \frac{\Delta \sqrt{N}\gamma_{\max}}{1-\gamma_{\max}}\,.
    \label{eq:error_bound}
\end{align}
\end{proposition}
\begin{proof}
For the sake of notational simplicity, define
$\mathbf{s}(t)=\E_{\phib} \E_\v \mathbf{e}(t)$. The dynamics of
$\mathbf{s}(t)$ are thus given by a deterministic time-varying linear
system. Consider the function
$V(\mathbf{s}(t))=\|\mathbf{s}(t)\|$. Simple algebra gives that
\begin{align*}
V(\mathbf{s}(t)) &\leq \|\E_{\phib} (\K(t) \circ \Phi(t))\|
V(\mathbf{s}(t-1)) + \|\E_{\phib} (\K(t) \circ \Phi(t))\| \Delta \sqrt{N} \,.
\end{align*}
Now, consider that
\begin{align*}
\E_{\phib} (\K(t) \circ \Phi(t)) = \sum_{\varphib_{|t} \in\: \Sigma} (\K(t) \circ \varphib_{|t}) \Pr(\varphib_{|t}) \,,
\end{align*}
where $\Pr(\varphib_{|t})$ is the probability of the packet reception realization $\varphib_{|t}$ at time $t$. The expectation is given by the sum of a finite number of combinations $\Sigma$ of possible packet loss realizations, since the network has a finite number of links, and in each link a packet can be either successfully received or dropped, so that $|\Sigma| = 2^{|\mathcal{E}|}$. It follows
\begin{align*}
\|\E_{\phib} \K(t) \circ \Phi(t) \| = \| \sum_{\varphib_{|t}\in\: \Sigma} (\K(t) \circ \varphib_{|t}) \Pr(\varphib_{|t}) \| \leq \sum_{\varphib_{|t}\in\: \Sigma} \|\K(t) \circ \varphib_{|t}\| \Pr(\varphib_{|t}) \leq \gamma_{\max}\,,
\end{align*}
because obviously $\sum_{\varphib_{|t}\in\: \Sigma}
\Pr(\varphib_{|t}) = 1$, and, from assumption (i), $\|\K(t) \circ
\varphib_{|t}\| \leq \gamma_{\max}<1$. Therefore
$$
    V(\mathbf{s}(t)) \leq \gamma_{\max}^t V(\mathbf{s}(t-1)) + \gamma_{\max}
    \frac{1-\gamma_{\max}^t}{1-\gamma_{\max}}\Delta\sqrt{N}\,,
$$
from where, taking the limit $t\rightarrow +\infty$, the proposition follows.
\end{proof}

\begin{remark}
Notice that the expected error converges to a neighborhood of the
origin exponentially fast, and more precisely with rate
$\gamma_{\max}<1$.
\end{remark}


Proposition~\ref{prop:error-stability} provides us with conditions
for the convergence of the estimation error of the entire network
to a neighborhood of the origin. It follows that the estimation
error of the entire network is subject to a cumulative bias. It is
clear that such a bias depends on $\|\K(t) \circ \varphib_{|t}\|$
and $\Delta$. If the signal $d(t)$ is slowly varying, namely
$\Delta\ll 1$, and $\|\K(t) \circ \varphib_{|t}\|$ small, then the
bias will be small.


Notice that, in order to ensure that the estimation error
decreases at each node, a condition at network level is
required, namely it must hold that
$\gamma(\K(t) \circ \varphib_{|t})\leq\gamma_{\max}<1$. 
The constant $\gamma_{\max}$
can be chosen by fixing a maximum cumulative estimation
error and solving~\eqref{eq:error_bound} for
$\gamma_{\max}$, as we show in Section~\ref{sec:simulations}.

We will show in the next subsection how to choose the
weights $\k_i(t)$ and $\h_i(t)$ locally at each node so that the estimation
error variance is minimized.

\subsection{Distributed Computation of Filter Coefficients}

To design a minimum variance distributed estimator, we
need to consider how the error variance evolves over time. The
estimation error variance dynamic at node $i$ is given by
\begin{align}
    \E_\v (e_i(t) - \E_\v e_i(t))^2 =
    (\k_i\phib_i)^T(t) \P_i(t-1) (\k_i\phib_i)(t)+
    \sigma^2 (\h_i\phib_i)^T(t) (\h_i\phib_i)(t)
    \label{eq:error_cov}
\end{align}
where
$$
    \P_i(t-1) = \E (\mathbf{e}_i(t) - \E \mathbf{e}_i(t))\E (\mathbf{e}_i(t) - \E
    \mathbf{e}_i(t))^T \in \R^{N \times N}\,.
$$
We assume that $\x_i(0) = \u_i(0)$, and
$$
    \e_i(0) = \u_i(0) - \frac{1}{|\mathcal{N}_{\varphib_i}| -1 } \mathop{\sum_{j=1}^{|\mathcal{N}_{\varphib_i}|}}_{j\neq i} \u_j(0) \,,
$$
because an initial (rough) estimate of $d(0)$ can be computed as
arithmetic average of the $|\mathcal{N}_{\varphib_i}| -1$
measurements received from neighboring nodes.

The estimation error variance~\eqref{eq:error_cov} can be
rewritten as
$$
    \E_\v (e_i(t) - \E_\v e_i(t))^2 = \k_i^T(t) \left(\P_i(t-1) \circ
    (\phib_i(t) \phib_i^T(t))\right) \k_i(t) + \sigma^2 \h_i^T(t)
     \phib_i(t) \phib_i^T(t)  \h_i^T(t)
$$
where we used the fact that $\mathbf{a}^T \mathbf{B}
\mathbf{a} = \trace (\mathbf{B} \mathbf{aa}^T)$ and $\trace
(\mathbf{A} \circ \mathbf{B}) \mathbf{C} = \trace
(\mathbf{C}^T \circ \mathbf{B}) \mathbf{A}^T$.

The optimal weights $\k_i(t)$ and $\h_i(t)$ are chosen so that at
each time instant, for any given realization of the packet loss
process $\varphib_i{|t}$ of $\phib_i(t)$, the variance $E_\v
(e_i(t) - \E_\v e_i(t)|\phib_i(t)=\varphib_{i|t})^2$ is minimized, under the constrain~\eqref{eq:cond_1} and that $\gamma(\K(t) \circ \varphib_{|t})\leq \gamma_{\max} <1$.
As already mentioned, the second constraint is global, since $\K(t) \circ \varphib_{|t}$
depends on all $\k_i(t)$, $i=1,\ldots,N$. However, it is possible to determine local conditions so that the global constraint is satisfied. We show next how this can be done.

For $i=1,\dots,N$, we define the set
$\Theta_{\varphib_i}=\{j\neq i : \mathcal{N}_{\varphib_i}
\cap \mathcal{N}_{\varphib_j} \neq \emptyset\} \cup
\{\mathcal{N}_{\varphib_i} \}$, which is the collection of
communicating nodes located at two hops distance from
node~$i$ plus communicating neighbors of~$i$, at time~$t$.
The following result holds:
\begin{proposition}
    Suppose there exists $\psib(t)=(\psi_1(t), \psi_2(t), \ldots, \psi_N(t) )^T \succ 0$, such that
    \begin{align}
        \psi_i(t) + \sqrt{\psi_i(t)} \sum_{j\in
        \Theta_{\varphib_i}}\sqrt{\psi_j(t)}
        \leq \gamma_{\max} \quad \forall i\,. \label{eq:part-proof}
    \end{align}
    If $\|\k_i(t)\circ \varphib_{i|t}\|^2 \leq \psi_i(t)$, $i=1,\dots,N$, then
    ${\gamma}(\K(t)\circ \varphib_{|t}) \leq \gamma_{\max} < 1$.
    \label{prop:gershorin-like}
\end{proposition}
\begin{proof}
The proof is similar to the proof of Proposition~III.1 in~\cite{JSAC07}.
\end{proof}
Using this proposition, the global constraint ${\gamma}(\K(t) \circ \varphib_{|t})\leq \gamma_{\max} <1$ can be replaced by the constraint $\|\k_i(t) \circ \varphib_{i|t}\|^2 \leq
\psi_i(t)$, where $\psi_i(t)$ satisfies the set of nonlinear
inequalities~\eqref{eq:part-proof}. Therefore, each node needs to
solve the following optimization problem
\begin{align}
    \min_{\k_i(t),\h_i(t)} \quad & \k_i^T(t) \left(\P_i(t-1) \circ
    (\varphib_{i|t} \varphib_{i|t}^T)\right)
    \k_i(t)+\sigma^2 \h_i^T(t) \varphib_{i|t} \varphib_{i|t}^T \h_i(t) \label{eq:local-optimiz-probl1}\\
    \text{s.t.}\quad & \left((\k_i(t)+\h_i(t))^T \circ \varphib_{i|t}\right)\1=1\nonumber\\
    &\|\k_i(t) \circ \varphib_{i|t}\|^2 \leq \psi_i(t)\nonumber\,.
\end{align}
We will discuss in Section~\ref{sec:computation_of_psi} how to
compute the values of $\psi_i(t)$, which are needed to state
problem~\eqref{eq:local-optimiz-probl1}. Observe that the
optimization problem~\eqref{eq:local-optimiz-probl1} is a
Quadratically Constrained Quadratic
Problem~\cite[pag.~653]{boyd2}. It admits a strict interior point
solution, corresponding to $\k_i(t)=0$ and $(\h_i(t) \circ
\varphib_{i|t})\:\1=1$. Thus Slater's condition is satisfied and
strong duality holds~\cite[pag.~226]{boyd2}. The problem, however,
does not have a closed form solution and thus we need to rely on
numerical algorithms to derive the optimal $\k_i(t)$ and
$\h_i(t)$. We have the following proposition.
\begin{proposition}
    \label{prop:optimal-values}
    For a given covariance matrix $\P_i(t-1)$ and a realization $\varphib_{i|t}$ of $\phib_i(t)$, the values of $\k_i(t)$ and $\h_i(t)$ that
    minimizes~\eqref{eq:local-optimiz-probl1} are
    \begin{align}
        \k_i(t) &= \frac{ \left((\P_i(t-1)+\lambda_i(t) \I) \circ \varphib_{i|t}\varphib_{i|t}^T\right)^\dag\varphib_{i|t}}
        {\varphib_{i|t}^T\left(\left((\P_i(t-1)+\lambda_i(t) \I) \circ \varphib_{i|t}\varphib_{i|t}^T\right)^\dag + \sigma^{-2}\I \right) \varphib_{i|t}} \,, \label{eq:gainsk} \\
        \h_i(t) &= \frac{\left(\varphib_{i|t} \varphib_{i|t}^T\right)^\dag\varphib_{i|t}}
        {\sigma^2\varphib_{i|t}^T  \left(\left((\P_i(t-1)+\lambda_i(t) \I) \circ \varphib_{i|t}\varphib_{i|t}^T\right)^\dag + \sigma^{-2}\I\right)\varphib_{i|t}} \,,
        \label{eq:gains}
    \end{align}
    with the optimal Lagrange multiplier $\lambda_i(t)\in  \left[0, \max\left(0,\sigma^2/\sqrt{|\mathcal{N}_{\varphib_i}|
    \psi_i(t)} - \ell_m (\Gammab_i(t-1))\right)\right]$.
\end{proposition}
\begin{proof}
The proof is similar to the proof of Proposition~III.2
in~\cite{JSAC07}.
\end{proof}
\begin{remark}
Modeling the packet loss by the Hadamard product allows us to
obtain weights having a similar form to those we obtained in the
case of no packet loss~\cite{JSAC07}. However, this result is not
a straightforward application of~\cite{JSAC07}
because~\eqref{eq:gainsk} and~\eqref{eq:gains} are obtained by
exploitation of the Hadamard product and the Moore-Penrose
pseudo-inverse in the computation of the Lagrange dual function
and the KKT conditions. Therefore, the previous proposition
generalizes our earlier result for any given realization of the
packet loss process. In the special case when $\varphib_{i|t} =
\1$, namely when there are no packet losses, we reobtain the
result in~\cite{JSAC07}.
\end{remark}

Previous proposition provides us with an interval within which the
optimal $\lambda_i$ is located. Simple search algorithms can be
considered to solve numerically $(\k_i)^T \k_i-\psi_i=0$ for
$\lambda_i$, such as, for example, the bisection algorithm.

It is worth noting that $k_{ij}$, and similarly $h_{ij}$,
in~\eqref{eq:gains} are zero if node~$j$ does not communicate with
node~$i$ because of a lost packet. In such a case the $j$-th
row and column of the matrix $\varphib_{i|t}\varphib_{i|t}^T$ is
zero. The pseudo-inverse $((\P_i(t-1)+\lambda_i(t) \I) \circ
\varphib_{i|t}\varphib_{i|t}^T)^\dag$ maintains the zeros in the
same position as those in the matrix
$\varphib_{i|t}\varphib_{i|t}^T$.

\subsection{Error Covariance Matrix}

Proposition~\ref{prop:optimal-values} provides us with the optimal
weights that minimize the estimation error variance at each time
step. The optimal weights $\k_i(t)$ and $\h_i(t)$ depend
indirectly on the thresholds $\psi_i(t)$, through the Lagrangian
multiplier $\lambda_i(t)$, and directly on the error covariance
matrix $\P_i(t-1)$. We will discuss in the next section how it is
possible to compute such thresholds $\psi_i(t)$ in a distributed
way, whereas we dedicate the rest of this subsection on discussing
how to locally compute the error covariance matrix $\P_i(t-1)$.
More precisely, because of the packet loss process, node~$i$
requires only the elements of $\P_i(t-1)$ corresponding to its
neighbors, namely the matrix $\P_i(t-1) \circ
\varphib_{i|t}\varphib_{i|t}^T$.

Each node can estimate from data the error covariance matrix,
which we denote with $\hat{\Gammab}_i(t)$, as discussed
in~\cite{JSAC07}. However, here we need to extend the approach to
the case of packet losses, because the design of the estimator of
the covariance matrix is tricky when packets are lost. If a node
$j$ exchanges data with its neighboring node~$i$, after an outage
period, node~$i$ needs to re-initialize the $j$-th row and column
of $\hat{\Gammab}_i(t)$ reasonably in order to take advantage of
the new acquired neighbor. We consider the following
re-initialization of elements of the error covariance matrix
$\hat{\Gammab}_i(t)$.

If at time~$t$ a new neighbor of a node is exchanging data, then
the diagonal element of the estimate of the error covariance
matrix at time $t-1$, corresponding to such a neighbor, is
initialized to the maximum element in the diagonal of the error
covariance matrix. More precisely, let  $j\in \mathcal{N}_i$ and
assume that for $t\in (t_1,t_2)$, $j \notin
\mathcal{N}_{\varphib_{i|t_1}}$, and that $j \in
\mathcal{N}_{\varphib_{i|t_2}}$. Then
\begin{align*}
    \left[\hat{\Gammab}_i(t_2-1)\right]_{jj} := \max_k \left[\hat{\Gammab}_i(t_2-1)\right]_{kk}
\end{align*}
and for $\ell \in \mathcal{N}_i$,
\begin{align*}
    \left[\hat{\Gammab}_i(t_2-1)\right]_{\ell j} & := \left[\hat{\Gammab}_i(t_2-1)\right]_{j \ell} := 0\,.
\end{align*}
This heuristic is motivated by the fact that all nodes are
collaborating to build and estimate of $d(t)$, and they are using
the same algorithm. Thus the maximum variance of the estimation
error that a neighbor of a node is affected by must not be larger
than the worst variance of the estimation error of other
neighbors. Obviously, chances are that the heuristic might
overestimate the variance associated to a new neighbor. However,
from simulations in Section~\ref{sec:simulations} we see that this
strategy works well in practice, even in the presence of high packet loss
probabilities.

\section{Computation of the Thresholds}
\label{sec:computation_of_psi}

From Proposition~\ref{prop:gershorin-like}, we notice that
thresholds $\psi_i$'s need to be upper bounded to guarantee
convergence of the estimation error. It holds that the larger the
value of $\psi_i(t)$ the lower the error variance. Indeed, after
some algebra, it follows that
\begin{align}
   \E_\v (e_i(t) - \E_\v e_i(t) | \phib_i(t)=\varphib_{i|t})^2
   \leq \frac{\sigma^2}{\sigma^2\varphib_{i|t}^T  \left(\left((\P(t-1)+\lambda_i(t) \I) \circ \varphib_{i|t}\varphib_{i|t}^T\right)^\dag + \sigma^{-2}
   \I  \right)\varphib_{i|t}}\,.
   \label{eq:optim_value_var}
\end{align}
From this inequality we have that the estimation error variance at
the node~$i$ decreases as $\lambda_i(t)$ decreases. From
Proposition~\ref{prop:optimal-values} we see that if $\psi_i$ is
large, then the Lagrangian multiplier $\lambda_i$ is small, since
$\lambda_i\in  \left[0, \max\left(0,
\sigma^2/\sqrt{|\mathcal{N}_{\varphib_i}| \psi_i} - \ell_m
(\Gammab_i)\right)\right]$.

According to the arguments above, we are interested in determining
the largest solution of the nonlinear equations in
Proposition~\ref{prop:gershorin-like}. Therefore, we consider the
following optimization problem:
\begin{align}
    \max_{\psib(t)}\qquad & \1^T\psib(t) \label{optp:maxpsi}  \\
    \mathrm{s.t.} \qquad & \S(\psib(t)) \preceq 0 \label{eq:constraint_psi} \\
    &\psib(t) \succ 0 \nonumber \,,
\end{align}
where $\S(\psib(t))=(S_1(\psib(t)), \ldots, S_N(\psib(t)))^T$ and
\begin{align*}
    S_i(\psib(t))=\psi_i(t) + \sqrt{\psi_i(t)}
    \mathop{\sum_{j\in \Theta_{\varphib_i}}}\sqrt{\psi_j(t)} - \gamma_{\max} \,.
\end{align*}
The solution of previous problem can be computed easily via
standard centralized approaches, but in our setup the computation
of the solution must be obtained in a decentralized fashion. The
distributed computation of the solution could be performed through
message passing, as in~\cite{JohannsonXiao06}. However, the
converge speed is prohibitive. Hence, we consider an alternative
approach. The fact that in~\eqref{eq:constraint_psi} only
information from two-hop neighboring nodes is required, and not of
the entire network, allows us to develop a decentralized algorithm
to compute the optimal solution. This is obtained in two steps.
First we show that the optimal solution satisfies the inequality
constraints~\eqref{eq:constraint_psi} with equality. Second, we build on this to distribute
the computation among nodes to obtain the optimal solution. We
provide details in the sequel.

\subsection{Equality constraints}

In this section, we show that there is a global optimal solution
of~\eqref{optp:maxpsi} that satisfies the inequality
constraints~\eqref{eq:constraint_psi} with equality. In particular
we have the following important result.

\begin{theorem} \label{theo:OP}
Problem~\eqref{optp:maxpsi} admits a global optimum $\psib^*(t)$, which is the solution of the following set of nonlinear equations:
\begin{align}
    \psi_i^*(t) +  \sqrt{\psi_i^*(t)} \mathop{\sum_{j\in
    \Theta_{\varphib_i}}} \sqrt{\psi_j^*(t)} - \gamma_{\max} = 0 \qquad i=1,\dots,N\,,\label{eq:eqconstr}
\end{align}
where $\Theta_{\varphib_i}=\{j\neq i : \mathcal{N}_{\varphib_i}
\cap \mathcal{N}_{\varphib_j} \neq \emptyset\} \cup
\{\mathcal{N}_{\varphib_i} \}$.
\end{theorem}

To prove this theorem, we need some intermediate technical results:

\begin{lemma}
\label{lemma:psi-bound}
There exists a feasible solution $\psib^\ell(t) =(\psi_1^\ell(t), \ldots, \psi_i^\ell(t), \ldots,
\psi_N^\ell(t))^T \succ 0$ of~\eqref{optp:maxpsi}, where
\begin{align}
    \label{eq:ai} \psi_i^\ell(t)=\frac{\gamma_{\max}}{4}\left( \sqrt{|\Theta_{\varphib_i}|^2 + 4} - |\Theta_{\varphib_i}| \right)^2 \quad i=1,\dots,N \,.
\end{align}
\end{lemma}
\begin{proof}
The $i$-th element of $\psib^\ell$, $\psi_i^\ell$, is constructed by considering the $i$th constraint, and imposing that the other variables $\psi_j$, for $j\in \Theta_{\varphib_i}$, assume
the largest value, which is~$\gamma_{\max}$:
\begin{align*}
    \psi_i +  \sqrt{\psi_i} |\Theta_{\varphib_i}|\sqrt{\gamma_{\max}} = \gamma_{\max} \,.
\end{align*}
By solving the equation for $\psi_i=\psi_i^\ell$ we obtain
\begin{align}
     \psi_i^\ell=\frac{\gamma_{\max}}{4}\left( \sqrt{|\Theta_{\varphib_i}|^2 + 4} - |\Theta_{\varphib_i}| \right)^2 \,.
\end{align}
The same procedure can be repeated $\forall i=1,\ldots,N$. The $\psi_i^\ell$ obtained are collected into a vector $\psib^\ell =(\psi_1^\ell, \ldots, \psi_i^\ell, \ldots,
\psi_N^\ell)^T$. Since
\begin{align*}
    \psi_i^\ell +  \sqrt{\psi_i^\ell} \mathop{\sum_{j\in
    \Theta_{\varphib_i}}} \psi_j^\ell \leq \psi_i^\ell +  \sqrt{\psi_i^\ell}\:|\Theta_{\varphib_i}|\sqrt{\gamma_{\max}} = \gamma_{\max} \,,
\end{align*}
$\psib^\ell$ is a feasible solution.
\end{proof}
This lemma is useful, because it allows us to establish the
existence of an optimal solution:
\begin{lemma} \label{theo:OPsolutions}
Problem~\eqref{optp:maxpsi} admits an optimal solution $\psib^*(t)$, which is the solution of the following set of nonlinear equations:
\begin{align*}
    \psi_i^*(t) +  \sqrt{\psi_i^*(t)} \mathop{\sum_{j\in
    \Theta_{\varphib_i}}} \sqrt{\psi_j^*(t)} - \gamma_{\max} = 0 \qquad i=1,\dots,N \,.
\end{align*}
\end{lemma}
\begin{proof}
The proof is based on a useful rewriting of the optimization
problem and by a {\it reductio ad absurdum} argument.

Let $y_i^2=\psi_i$ for $i=1,\ldots,N$. Then, the optimization problem~\eqref{optp:maxpsi} can be rewritten as follows
\begin{align}
    \max_{\y}\qquad & \y^T\y \label{optp:maxy}  \\
    \mathrm{s.t.} \qquad & \y-\f(\y) \preceq 0 \label{eq:constrainty} \\
    &\y \succ 0 \nonumber \,.
\end{align}
where $\f(\y)=(f_1(\y), \ldots, f_N(\y))^T$ and
\begin{align*}
   f_i(\y)=y_i-\beta\left(y_i^2 + y_i\mathop{\sum_{j\in \Theta_{\varphib_i}}}y_j - \gamma_{\max}\right) \,,
\end{align*}
with $\beta$ being any positive scalar. This problem
and~\eqref{optp:maxpsi} are obviously equivalent: for all
$\beta>0$, $\S(\psib) \preceq 0$ if and only if $\y-\f(\y)\preceq
0$. Let $\y^*$ be an optimal solution of~\eqref{optp:maxy}, then
$\psi_i^*={y_i^*}^2$. Problem~\eqref{optp:maxy} admits optimal
solutions, since from Proposition~\ref{lemma:psi-bound} the
problem is feasible. We show next that the optimal solutions
satisfy the constraints at the equality.

Let $\y^*$ be an optimal solution. Suppose by contradiction that
there is constraint $i$ that is satisfied at a strict inequality,
namely $y_i^* < f_i(\y^*)$, while suppose $y_j^*\leq f_j(\y^*)$
for $i\neq j$. In the following, we show that from $\y^*$ we can
construct a feasible solution $\t^*$ such that ${\t^*}^T\t^* >
{\y^*}^T\y^* $, so that it is not possible that $\y^*$ be an
optimal solution.

Since $\beta$ is arbitrary, we can select a convenient value. Let
$$
\beta \leq \bar{\beta} < \min_{0 \prec \y \preceq 1} \frac{1}{2y_i+\mathop{\sum_{j\in \Theta_{\varphib_i}}}y_j} =\frac{1}{2+\Theta_{\varphib_i}} \,.
$$
This choice of $\beta$ makes $f_i(\y)$ being an increasing
function of $y_i$, and a decreasing function of $y_j$, for $j\neq
i$. Indeed
\begin{align*}
\nabla_i f_i(\y^*) &= 1-\beta\left(2y_i^* + \mathop{\sum_{j\in \Theta_{\varphib_i}}}y_j^*\right) > 0 \,,\\
\nabla_j f_i(\y^*)&= \left\{
                    \begin{array}{ll}
                      -\beta y_i^*  < 0, & \mbox{ \rm if } j\in \Theta_{\varphib_i} \\
                      0, & \mbox{ \rm if } j\notin \Theta_{\varphib_i}, j\neq i
                    \end{array}
                  \right.\\
\intertext{and}
\nabla^2_i f_i(\y^*)&= -2\beta <0 \,,\\
\nabla^2_j f_i(\y^*)&= 0 \quad  \mbox{ \rm if } \quad j\neq i \,,
\end{align*}
Let $\z\in \R^N$ such that $z_i \in (0,1]$. We have
\begin{align} \label{eq:taylor}
f_i(\z)=f_i(\y^*)+\nabla f_i(\y^*)(\z-\y^*)^T+ \frac{1}{2}(\z-\y^*)^T \nabla^2 f_i(\y^*)(\z-\y^*)\,,
\end{align}
because the third order derivatives are zero. Then, we chose a small positive scalar $0<\varepsilon \leq f_i(\y^*)-\y^*$ so that $\z$ be an augmented vector of $\y^*$, with $z_i=\varepsilon+y_i^*$, $z_j=y_j^*$ for $j=1,\ldots,N$, $j\neq i$, and $z_i =y_i^*+\varepsilon \leq f_i(\y^*) < f_i(\z)$. The last inequality is allowed by the fact that $f_i(\y)$ is an increasing function of $y_i$. From~\eqref{eq:taylor} it follows
\begin{align*}
f_i(\z) & = f_i(\y^*)+\nabla_i f_i(\y^*)\varepsilon+  \frac{1}{2}\varepsilon^2 \nabla_i^2 f_i(\y^*) = f_i(\y^*)+\left[1-\beta\left(2y_i^* + \mathop{\sum_{j\in \Theta_{\varphib_i}}}y_j^*\right) \right]\varepsilon-\beta\varepsilon^2 \\
& \triangleq f_i(\y^*) +\Delta f_i\,, \\
f_j(\z) & = f_j(\y^*)+\nabla_i f_j(\y^*)\varepsilon+  \frac{1}{2}\varepsilon^2 \nabla_i^2 f_j(\y^*) = f_j(\y^*)-\beta y_j^*\varepsilon\\
&  \triangleq f_j(\y^*) -\Delta f_j  \quad \mbox{\rm if }\quad j\in \Theta_{\varphib_i} \,, \\
f_\ell(\z) & = f_\ell(\y^*)  \quad \mbox{\rm otherwise } \,.
\end{align*}
By using $\varepsilon$ and $\Delta f_j$, $j\neq i$, we can define
a vector $\t^*$ such that $t_i^*=y_i^*+\varepsilon$,
$t_j^*=y_j^*-\Delta f_j$ if $j\in \Theta_{\varphib_i}$, and
$t_\ell^*=y_\ell^*$ otherwise. Notice that $\f(\z)\preceq\f(\t^*)$ since
$\t^*\preceq \z$. The solution $\t^*$ is feasible for
problem~\eqref{optp:maxy}, namely $\t^*\preceq \f(\t^*)$, because
$t_i^*=y_i^*+\varepsilon = z_i \leq f_i(\z) \leq f_i(\t^*)$,
$t_j^*=y_j^*-\Delta f_j \leq  f_j(\y^*) -\Delta f_j= f_j(\z) \leq
f_j(\t^*)$ if $j \in \Theta_{\varphib_i}$ and $t_\ell^* =y_\ell^* \leq
f_\ell(\y^*) = f_\ell(\t^*)$ if $\ell\neq i$ and $l \notin
\Theta_{\varphib_i}$. Now, observe that
\begin{align*}
{\t^*}^T\t^* - {\y^*}^T\y^* & =  \varepsilon^2 + \sum_{j\in \Theta_{\varphib_i}}^N \Delta f_j^2 + 2 y_i^*\varepsilon+
2 \sum_{j\in \Theta_{\varphib_i}}^N y_j^*\Delta f_j \\
& =  \varepsilon^2 +\beta^2 \varepsilon^2 \sum_{j\in
\Theta_{\varphib_i}}{y_j^*}^2 + 2y_i^* \varepsilon -2 \beta
\varepsilon  \sum_{j\in \Theta_{\varphib_i}} {y_j^*}^2\,.
\end{align*}
The last right-hand side of previous equation is always positive, provided that one chooses
\begin{align*}
\varepsilon < \frac{2\beta  \sum_{j\in \Theta_{\varphib_i}} {y_j^*}^2}{1+\beta^2 \sum_{j\in \Theta_{\varphib_i}}{y_j^*}^2}\,.
\end{align*}
This implies that ${\t^*}^T\t^* > {\y^*}^T\y^*$, namely that $\t^*$ is a feasible solution of~\eqref{optp:maxy} with higher cost function than $\y^*$, which is a contradiction because $\y^*$ was assumed to be an optimal solution. It follows that optimal solutions must satisfy all the constraints at the equality.
\end{proof}
The previous lemma guarantees that there are optimal solutions
satisfying the constraints at the equality. However, we do not
know yet if there is a global optimal solution. If there were
multiple optimal solutions, we would have to chose the most fair
for all nodes. Recall that a small $\psi_i^*(t)$ means smaller
estimation quality. To establish the uniqueness of the optimal
solution, we need the following lemma, which will be used for the
proof of Theorem~\ref{theo:OP}:

\begin{lemma} \label{lemma:Jacob}
Let $\J(\psib(t))=\nabla \S(\psib(t))$ be the Jacobian of $\S(\psib(t))$. Then $\J(\psib(t))$ is a nonsingular matrix.
\end{lemma}
\begin{proof}
The diagonal elements of the Jacobian are
$$
J_{ii}(\psib)= \nabla_i S_i(\psib) = 1+\frac{1}{2\sqrt{\psi_i}}\sum_{j\in \Theta_{\varphib_i}}\sqrt{\psi_j}\,,
$$
whereas the off-diagonal elements $\nabla_i S_j(\psib)$ are either
zero if $j \notin \Theta_{\varphib_i}$, or
$$
J_{ji}(\psib)=\nabla_i S_j(\psib)=\frac{\sqrt{\psi_j}}{2\sqrt{\psi_i}}\quad \mbox{ if } j\in \Theta_{\varphib_i} \,.
$$
By applying the Gershgorin theorem, we have that the eigenvalues of the Jacobian lie in the region
\begin{align*}
    \ell_k(\mathbf{J}(\psib)) \in \bigcup_{i=1}^N \big{\{}z \in \C : |z-J_{ii}(\psib) | \leq \sum_{j \neq i } |J_{ji}|\big \} \quad \forall k\,,
\end{align*}
from which it follows that the real part of the minimum eigenvalue is such that
\begin{align*}
    {\rm Re }\: \ell_m(\J(\psib)) & \geq J_{ii}(\psib) - \sum_{j \neq i } |J_{ji}|  = 1+\frac{1}{2\sqrt{\psi_i}}\sum_{j\in \Theta_{\varphib_i}}\sqrt{\psi_j} - {\sum_{j\in \Theta_{\varphib_i}}} \frac{1}{2\sqrt{\psi_i}} \sqrt{\psi_j} = 1 >0 \quad \forall i\,.
\end{align*}
Therefore, $\J(\psib)$ has no zero eigenvalues, namely it is non-singular.
\end{proof}

We are now in the position of proving Theorem~\ref{theo:OP}. From
Lemma~\ref{theo:OPsolutions}, we know that there is an optimal
solution satisfying the constraints at the equality. We show next
that such a solution is unique, thus proving
Theorem~\ref{theo:OP}.
\begin{proof}[Proof of Theorem~\ref{theo:OP}]
The proof of the uniqueness of the optimal solution is based on
the use of the Lagrange dual theory. First, observe that from
Lemma~\ref{theo:OPsolutions} the optimization problem admits
optimal solutions. The optimization problem is non-convex, since
the constraints~\eqref{eq:constraint_psi} are not convex. The
Lagrange dual theory for non-convex non-linear optimization
problems can be applied. A qualification constraint
from~\cite[pag. 25]{Horst+} states that strong duality holds if
the optimization problem is feasible and the Jacobian of
$\S(\psib(t))$ is non-singular, which we know from
Lemma~\ref{lemma:psi-bound} and Lemma~\ref{lemma:Jacob},
respectively. Therefore, the optimal solution of the problem can
be investigated via the Lagrange dual function $L(\xib,\psib) = -
\psib^T \1  + \xib^T \S(\psib)$, where $\xib\succeq \0$ is the
Lagrangian multiplier. From the KKT conditions it follows that
$\J(\psib)^T \xib  = \1 $. We see that previous equality trivially
holds also for the optimal solution $\psib^*$, namely
$\J(\psib^*)^T \xib^* = \1$. Since from Lemma~\ref{lemma:Jacob}
the Jacobian is non-singular, it follows that there is a unique
solution to the previous system of equations, namely
$\xib^*=\J(\psib^*)^{-T}\1$, and since strong duality holds, we
conclude that the optimal solution given by~\eqref{eq:eqconstr} is
unique.
\end{proof}

\begin{corollary} \label{cor:lower-bound-opt-sol}
Let $\psib^*(t)$ be the solution of~\ref{eq:eqconstr}. Then,
$\psib^*(t) \succeq \psib^\ell(t)$, where $\psib^\ell(t)$ is given
by~\eqref{eq:ai}.
\end{corollary}
\begin{proof}
The simple proof is by contradiction. Suppose that $\psib^\ell$ is not a lower bound on the optimal solution $\psib^*$, namely there is some $i$ for which $\psi_i^* < \psi_i^\ell$. By observing that that $\psib^* \preceq \gamma_{\max}\1$, it follows
$$\psi_i^* +  \sqrt{\psi_i^*} \cdot
    \mathop{\sum_{j\in \Theta_{\varphib_i}}}\sqrt{\psi_j^*} < \psi_i^\ell+ \sqrt{\psi_i^\ell} \cdot
    \mathop{\sum_{j\in \Theta_{\varphib_i}}}\sqrt{\psi_j^*} \leq  \psi_i^\ell+ \sqrt{\psi_i^\ell} |\Theta_{\varphib_i}|\sqrt{\gamma_{\max}} \leq \gamma_{\max} \,,
$$
which is a contradiction, because the optimal solution must satisfy the constraints of~\ref{optp:maxpsi} at the equality.
\end{proof}

We use Theorem~\ref{theo:OP} and
Corollary~\ref{cor:lower-bound-opt-sol} in the next sections to
develop a strategy for the distributed computation of the optimal
solution.

\subsection{Distribution of the Computation}

From the previous section, we compute the thresholds to use
in~\eqref{eq:local-optimiz-probl1} by the system of nonlinear
equations~\eqref{eq:eqconstr}. Unfortunately, an explicit solution
for such a system is not available. Numerical techniques have to
be used. In the following, we present a quick decentralized
algorithm with certified convergence.

We define the class of functions parameterized in the scalar
$\rho_i$
\begin{align}
    \label{eq:mapping2}
    \Rb_i(\psib)=\psi_i - \rho_i f_i(\psib) \,,
\end{align}
where $ f_i(\psib)= \psi_i +  \sqrt{\psi_i} \cdot \mathop{\sum_{j\in \Theta_{\varphib_i}}}\sqrt{\psi_j} - \gamma_{\max}$.
When $\Rb(\psib)$ is contractive, then it is easy to show that the fixed point of the mapping is the solution of~\eqref{eq:eqconstr}~\cite[Pag.191]{tsi}.  Furthermore, the convergence speed can be tuned at a local node $i$ by the parameter $\rho_i$. We have the following result
\begin{proposition} \label{prop:fastcontrmapping}
Let
\begin{align} \label{eq:gammat}
\rho^*_i(k)= \left\{
             \begin{array}{ll}
               \cfrac{2\sqrt{\psi_i(k)}}{{2\sqrt{\psi_i(k)}+\mathop{\sum_{j\in \Theta_{\varphib_i}}} \sqrt{\psi_j(k)}}}  & \mbox{ \rm if } 1 +  \cfrac{1}{2\sqrt{\psi_i(k)}} \mathop{\sum_{j\in \Theta_{\varphib_i}}}\sqrt{\psi_j(k)} \geq \mathop{\sum_{j\in \Theta_{\varphib_i}}}\cfrac{\sqrt{\psi_i(k)} }{2\sqrt{\psi_j(k)}} \\
               0 & \mbox{\rm otherwise\,.}
             \end{array}
           \right.
\end{align}
Then $\psi_i(k+1)=R_i(\psib(k))= \psi_i(k) - \rho_i(k) f_i(\psib(k))$ is a contraction mapping having the largest convergence
speed among the mappings~\eqref{eq:mapping2}.
\end{proposition}

\begin{proof}
Proposition~$1.10$ in~\cite[Pag.193]{tsi} gives a sufficient
condition to establish that~\eqref{eq:mapping2} is a contraction
mapping. If
\begin{align*}
1 > \alpha_i(k)& =|1-\rho_i(k)\nabla_i f_i(\psib(k)) |+
\sum_{j\neq i} |
\rho_i(k) \nabla_j f_i(\psib(k))| \\
& = \left|1-\rho_i(k)\left(1 +  \frac{1}{2\sqrt{\psi_i(k)}} \cdot \mathop{\sum_{j\in \Theta_{\varphib_i}}}\sqrt{\psi_j(k)}\right) \right| + \rho_i(k)\mathop{\sum_{j\in \Theta_{\varphib_i}}}\frac{\sqrt{\psi_i(k)}}{2\sqrt{\psi_j(k)}}  \,,
\end{align*}
where $\nabla_j$ is partial derivative operator with respect to
$\psi_j(k)$, then~\eqref{eq:mapping2} is contractive. The scalar
$\alpha_i(t)$ determines the converge speed of the mapping, so
that the lower is $\alpha_i(k)$ the faster is the convergence.

Suppose that
\begin{align} \label{eq:rho}
1 +  \frac{1}{2\sqrt{\psi_i(k)}} \cdot \mathop{\sum_{j\in \Theta_{\varphib_i}}}\sqrt{\psi_j(k)} \geq \mathop{\sum_{j\in \Theta_{\varphib_i}}}\frac{\sqrt{\psi_i(k)}}{2\sqrt{\psi_j(k)}} \,.
\end{align}
Then, $\alpha_i(k)$ is minimized if
$$
\rho_i(k) = \frac{1}{1 + \frac{1}{2\sqrt{\psi_i(k)}} \cdot \mathop{\sum_{j\in \Theta_{\varphib_i}}}\sqrt{\psi_j(k)} }  \,.
$$
Suppose that~\eqref{eq:rho} does not hold, than $\alpha_i(k)$ is minimized if $\rho_i(k)=0$. By putting together these cases, the proposition follows.
\end{proof}
From previous proposition, the overall mapping $\psib=\Rb(\psib)$,
where $\Rb:\R^N_+\rightarrow \R^N_+$, is a contraction mapping. The component solution method~\cite[Pag.187]{tsi} can be applied.
The solution of~\eqref{optp:maxpsi} is given by the algorithm
\begin{align}
    \psib(k+1) = \Rb(\psib(k))=(R_1(\psib(k)),\dots,R_N(\psib(k)))^T \,.
    \label{eq:iterative_algo3}
\end{align}
Using the $\rho^*_i(k)$ given by Proposition~\ref{prop:fastcontrmapping}, the mapping converges quickly. From Monte Carlo simulations, we see that the algorithm converges in less than 10 iterations on average.

\subsection{Algorithm for the Computation of the Thresholds}

The distributed computation of the thresholds~$\psi_i$ requires that the
neighboring nodes communicate the instantaneous values of the
local threshold, until~\eqref{eq:iterative_algo3} converges.
Clearly, the thresholds are over the same wireless channel used
for broadcasting estimates and measurements, and thus they are
subject to packet losses. These losses may happen during the phase
between the beginning of the iterations~\eqref{eq:iterative_algo3}
and the convergence. As a result, no convergence may be reached.
In the following, we develop a strategy to cope with this problem.


First, notice that the optimization problem is not sensitive to
perturbations of the constraints. In other words, if $\psib^*(t)$
is the solution of the system of non linear
equations~\eqref{eq:eqconstr}, then $\psib^*(t)$ is not
significantly perturbed by packet losses. We can see this from the
proof of Theorem~\ref{theo:OP}, form where we know that the
optimal solution is such that $\J(\psib^*)^T \xib^* = \1 $, with
$\J(\psib^*)$ being the Jacobian of the constraints and $\xib^*$
the Lagrange multipliers associated to the dual problem
of~\eqref{optp:maxpsi}. Specifically, the $i$-th equation of
$\J(\psib^*)^T \xib^*  = \1 $ is given by
\begin{align} \label{eq:lagrangefunc}
     \xi_i^*\left(1+\frac{1}{2\sqrt{\psi_i^*}}\sum_{j\in \Theta_{\varphib_i}}\sqrt{\psi_j^*}\right)
     + \sum_{j\in \Theta_{\varphib_i}} \xi_j^* \frac{\sqrt{\psi_j^*}}{2\sqrt{\psi_i^*}}= 1 \quad i=1,\dots,N \,.
\end{align}
This system of equations has positive coefficients, and
$\left(1+1/(2\sqrt{\psi_i^*})\sum_{j\in
\Theta_{\varphib_i}}\sqrt{\psi_j^*}\right) > 1$. Since $\xib^*
\succeq 0$, for strong duality holds, it follows that $\xi_i^* <1$
for $i=1,\ldots,N$. Then, $\xib^*<1$ implies that the optimal
solution is not sensitive to perturbations of the
constraints~\cite[pag. 249]{boyd2}.

Since a change in the number of two-hops neighbors of a node,
caused by packet losses, can be regarded as a perturbation of the
constraints, we conclude that the optimal solution of the
problem~\eqref{optp:maxpsi} is not much sensitive to the packet
losses. By this argument, we can compute just once the optimal
solution. In particular, we assume that the nodes compute the
optimal thresholds before the estimation algorithm starts by
considering the maximum number of neighbors. This is accomplished
by using high transmission radio powers and a retransmission
protocol that guarantee a successful packet reception. Such a
preliminary phase is very short, since from
Proposition~\ref{prop:fastcontrmapping} the computation of the
thresholds according to~\eqref{eq:iterative_algo3} requires few
iterations to converge.
During the estimation phase, if the packet loss probability is
very high, the perturbation might be large, resulting in a
significant change of the optimum. However, simulations reported
in Section~\ref{sec:simulations} show that the solution we adopt
for the threshold computation is robust to rather intense packet
losses.

\section{Performance Analysis}
\label{sec:performance_analysis}

In this section we characterize the performance of our estimator
by investigating the variance of the estimation error. We have the
following results:
\begin{proposition}
    \label{cor:bound-variance-node-i}
    For any packet loss realization $\varphib_{i|t}$ of $\phib_i(t)$, the optimal value of $\k_i(t)$ and $\h_i(t)$ are such that the error variance at node~$i$ satisfies
    $$
        \E_v (e_i^2-\E e_i^2|\phib_i(t)=\varphib_{i|t})^2 < \frac{\sigma^2}{|\mathcal{N}_{\varphib_i}|}\,.
    $$
\end{proposition}

\begin{proof}
From~\eqref{eq:optim_value_var} the error variance is upper-bounded
by
\begin{align*}
   \E_\v (e_i(t) - \E_\v e_i(t) | \phib_i(t) = \varphib_{i|t})^2 & \leq  \frac{\sigma^2}{\sigma^2\varphib_{i|t}^T \left((\P(t-1)+\lambda_i(t) \I) \circ \varphib_{i|t}\varphib_{i|t}^T\right)^\dag\varphib_{i|t} + \varphib_{i|t}^T \varphib_{i|t} } < \frac{\sigma^2}{ |\mathcal{N}_{\varphib_i}|}\,,
\end{align*}
where the inequality comes from the fact that $\varphib_{i|t}^T \left((\P(t-1)+\lambda_i(t) \I) \circ \varphib_{i|t}\varphib_{i|t}^T\right)^\dag\varphib_{i|t} > 0$, and recalling that $\varphib_{i|t}^T \varphib_{i|t} = |\mathcal{N}_{\varphib_i}|$.
\end{proof}
Notice that previous proposition guarantees that the
estimation error at each time $t$, and in each node, is
always upper-bounded by the variance of the estimator that
just takes the averages of the received $u_i(t)$.

\begin{proposition}
    \label{prop:bound-lM-Gamma}
    For any packet loss realization $\varphib_{|t}$ of $\phib(t)$,
    $$
        \ell_M \left((\P(t-1)+\lambda_i(t) \I) \circ \varphib_{i|t}\varphib_{i|t}^T \right) < \sigma^2 \left(1+\frac{2N}{(\sqrt{5}-1)\sqrt{\gamma_{\max}}}\right) \,.
    $$
\end{proposition}

\begin{proof}
It holds that $\ell_M \left((\P(t-1)+\lambda_i(t) \I) \circ \varphib_{i|t}\varphib_{i|t}^T \right)= \ell_M \left((\P(t-1)\circ \varphib_{i|t}\varphib_{i|t}^T \right) +\lambda_i(t)$.
Recalling that $\lambda_i(t)\in  \left[0, \max\left(0, \sigma^2/\sqrt{|\mathcal{N}_{\varphib_i}|
    \psi_i(t)} - \ell_m (\Gammab_i(t-1))\right)\right]$ it follows
 $$
        \lambda_i(t) < \frac{\sigma^2}{\sqrt{|\mathcal{N}_{\varphib_i}| \psi_i(t)}} \leq  \frac{\sigma^2}{\sqrt{|\mathcal{N}_{\varphib_i}| \frac{\gamma_{\max}}{4}\left( \sqrt{|\Theta_{\varphib_i}|^2 + 4} - |\Theta_{\varphib_i}| \right)^2}} \leq \sigma^2\frac{2N}{(\sqrt{5}-1)\sqrt{\gamma_{\max}}} \,,
    $$
where previous inequality comes from Corollary~\ref{cor:lower-bound-opt-sol}, observing that $\left( \sqrt{|\Theta_{\varphib_i}|^2 + 4} - |\Theta_{\varphib_i}| \right) \geq (\sqrt{5}-1)/|\Theta_{\varphib_i}| \geq (\sqrt{5}-1)/N$, and that $|\mathcal{N}_{\varphib_i}| \geq 1$. Furthermore, $\ell_M \left((\P(t-1)\circ \varphib_i(t)\varphib_i^T(t) \right)$ can be upper-bounded using the Ger\v{s}hgorin theorem, and recalling that the diagonal elements of $\P(t-1)\circ \varphib_{i|t}\varphib_{i|t}^T$ are less than $\sigma^2/ |\mathcal{N}_{\varphib_i}|$ from Proposition~\ref{cor:bound-variance-node-i}, and that each diagonal element of a covariance matrix assumes the largest value along its row. Hence, it follows that $\ell_M \left((\P(t-1)\circ \varphib_{i|t}\varphib_{i|t}^T \right) \leq \sigma^2$.
Putting together previous inequality, and the upper bound on $\lambda_i(t)$, the proposition follows.

\end{proof}
\begin{lemma} \label{lemma:boundexpt}
\begin{align}
    \E_{\phib} [\phib_i^T\phib_i]^{-1} = \sum_{k = 0}^{|\mathcal{N}_{i}|-1} \frac{\chi(k)}{k+1} \,,
\end{align}
where
\begin{align}
    \chi(k) = \sum_{\ell = 1}^{|\mathcal{N}_{i}| - 1 \choose k} \left(\prod_{n = 1}^k q_{is(n)}
     \cdot \prod_{m = k+1}^{|\mathcal{N}_{i}| - 1} p_{is(m)}\right) \,,
    \label{eq:poly_coeff}
\end{align}
and the function $s:\{1,2,\dots,|\mathcal{N}_{i}| - 1\}
\rightarrow \{1,2,\dots,|\mathcal{N}_{i}| - 1\}$ is a permutation.
Namely the $k$-th coefficient of the polynomial is the sum of
$|\mathcal{N}_{i}| - 1 \choose k$ terms in which there are $k$
factors $q_{ij}$ and $|\mathcal{N}_{i}|- 1 - k$ factors $p_{ir}$
with $j\neq r$.
\end{lemma}
\begin{proof}
The random variable $\phib_i^T\phib_i=\sum_{j=1}^{N}\phi_{ij} = 1+
\sum_{j=1, j\neq i}^{|\mathcal{N}_{i}|}\phi_{ij} $ is given by the
sum of $|\mathcal{N}_{i}|-1$ independent Bernoulli random
variables having different parameter. Then, we have~\cite{Chao72}
\begin{align} \label{eq:momentgen}
    \E_{\phib} \left[\phib_i^T\phib_i\right]^{-1} = \int_0^1 g_1(z) dz \,,
\end{align}
where $g_1(z)$ is the probability generating function of $\phib_i^T \phib_i$:
\begin{align*}
    g_1(z)= \E z^{\phib_i^T \phib_i - 1} = \mathop{\prod_{j=1}}_{j\neq i}^{|\mathcal{N}_{i}|} \E z^{\phi_{ij}} =
     \mathop{\prod_{j=1}}_{j\neq i}^{|\mathcal{N}_{i}|} (q_{ij}+p_{ij}z) = \chi_0 + \chi_1 z + \dots + \chi_{|\mathcal{N}_{i}|-1} z^{|\mathcal{N}_{i}|-1}
\end{align*}
where the last equality is achieved by developing the product of
terms $q_{ij}+p_{ij}z$ in a polynomial in the general form. After
tedious manipulations, we see that the coefficients of the
polynomial are given by~\eqref{eq:poly_coeff}.
%
%
By using $g_1(z)$ in the integral~\eqref{eq:momentgen}, we obtain
the result.
%
\end{proof}
\begin{proposition}
For any packet loss realization $\varphib_{i|t}$ of $\phib_i(t)$,
it holds
    \begin{align} \label{eq:expected_value_var-slightly-bounded}
        \E_{\phib} \E_v (e_i^2-\E_v e_i^2)\leq \frac{(\sqrt{5}-1)\sqrt{\gamma_{\max}}+2N}{2(\sqrt{5}-1)\sqrt{\gamma_{\max}}
        +2N}\cdot \sum_{k = 0}^{|\mathcal{N}_{i}|-1} \frac{\chi(k)}{k+1}\: \sigma^2\,.
    \end{align}
\end{proposition}
\begin{proof}
    The variance at node~$i$ is bounded as in~\eqref{eq:optim_value_var}. Following the same steps as in the proof of
    Proposition~\ref{cor:bound-variance-node-i}, we have
    \begin{align} \label{eq:EPhiEvei2b}
        \E_{\phib} \E_v (e_i^2-\E_v e_i^2) \leq \E_{\phib} \frac{\sigma^2}{\sigma^2\phib_i^T \left((\P(t-1)+\lambda_i(t) \I) \circ \phib_i\phib_i^T\right)^\dag\phib_i + \phib_i^T \phib_i }\,.
    \end{align}
    This inequality is based on the fact that the argument of the statistical expectation is always positive and the expectation is taken over a positive distribution, thus the sign of the argument is maintained~\cite[pag.392]{Horn85}.
    From Proposition~\ref{prop:bound-lM-Gamma}, it
    follows
    $$
        \phib_i^T \left((\P(t-1)+\lambda_i(t) \I) \circ \phib_i\phib_i^T\right)^\dag\phib_i \geq
        \cfrac{\phib_i^T\phib_i}{\sigma^2 \left(1+\cfrac{2N}{(\sqrt{5}-1)\sqrt{\gamma_{\max}}}\right)} \,.
    $$
    By using previous inequality in \eqref{eq:EPhiEvei2b}, we have
    $$
        \E_{\phib} \E_v (e_i^2-\E_v e_i^2) \leq \frac{(\sqrt{5}-1)\sqrt{\gamma_{\max}}+2N}{2(\sqrt{5}-1)\sqrt{\gamma_{\max}}+2N}  \E_{\phib}
        \frac{\sigma^2}{\phib_i^T\phib_i}\,.
    $$
    The proposition follows by invoking Lemma~\ref{lemma:boundexpt}.
\end{proof}

\begin{figure}
    \centering
    \psfrag{x}[][]{$\gamma_{\max}$}
    \psfrag{b}[][]{$N$}
    \psfrag{t}[][]{}
    \psfrag{y}[b][]{$\cfrac{(\sqrt{5}-1)\sqrt{\gamma_{\max}}+2N}{2(\sqrt{5}-1)\sqrt{\gamma_{\max}}+2N}$}
    \includegraphics[width=0.55\hsize]{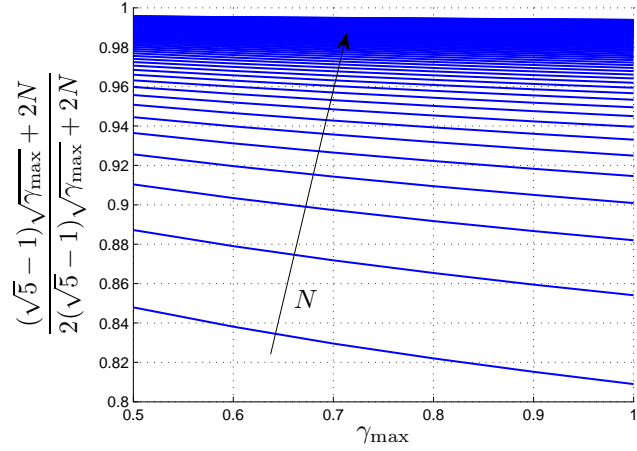}
    \caption{First factor of~\eqref{eq:expected_value_var-slightly-bounded} as function of $\gamma_{\max}$ for increasing values of $N$ ranging from 2 to 100.
    The factor is always less than 1 for all the values of the parameters. The smallest values are achieved when $\gamma_{\max}$ is large and $N$ is small.
    }\label{fig:e_phi_e_v_q}
\end{figure}

\begin{figure}
    \centering
    \psfrag{q}[][]{$q$}
    \psfrag{Ni}[l][]{$|\mathcal{N}_i|$}
    \psfrag{t}[][]{}
    \psfrag{y}[b][]{$\cfrac{1-q^{|\mathcal{N}_i|}}{(1-q)|\mathcal{N}_i|}$}
    \includegraphics[width=0.50\hsize]{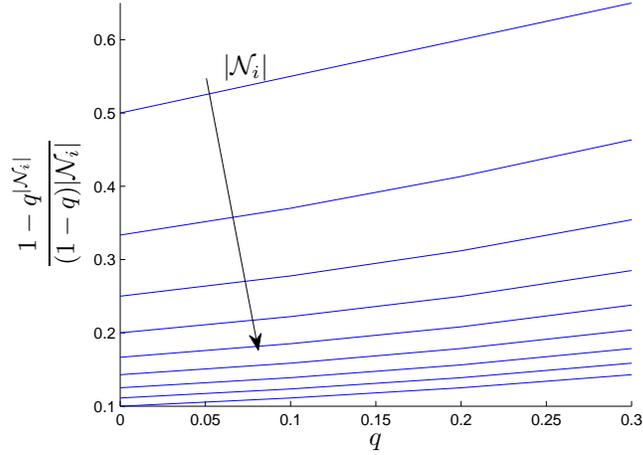}
    \caption{Second factor of~\eqref{eq:expected_value_var-slightly-bounded} as function of $q$ for increasing values of $|\mathcal{N}_i|$
    ranging from 1 to 20. The factor is always less than 1. The smallest values are achieved when
    $q$ is small and $|\mathcal{N}_i|$ is large. This is explained by the fact that the packet loss probability has a decreasing negative effect
    when the number of neighbors of a node increases, which translates into a smaller value of the coefficient.}\label{fig:coeff1}
\end{figure}

Observe that the estimation error variance given by the previous
proposition depends on the packet loss probabilities $q_{ij}$, on
the maximum number of neighbors for each node $|\mathcal{N}_{i}|$,
the total number of nodes in the networks $N$, and the largest
singular value of the matrix $\K(t)$. In
Figure~\ref{fig:e_phi_e_v_q} we have plotted the first factor of
the coefficient of~\eqref{eq:expected_value_var-slightly-bounded}.
It turns out that it is always less than 1. The smallest values
are achieved when $\gamma_{\max}$ is large and~$N$ small. The
second factor in~\eqref{eq:expected_value_var-slightly-bounded}
clearly depends on the value attained by the various $q_{ij}$. We
consider here the simple case when $q_{ij}=q$ for all~$i,j$, which
allows us writing the equations in closed form:
\begin{align}
    \sum_{k = 0}^{|\mathcal{N}_{i}|-1} \frac{\chi(k)}{k+1} = \frac{1-q^{|\mathcal{N}_{i}|}}{(1-q)|\mathcal{N}_{i}|}\,.
    \label{eq:coeff1_unif}
\end{align}
In Figure~\ref{fig:coeff1} we have plotted such a function for
various values of~$q$ and $|\mathcal{N}_{i}|$. The function
decreases very fast as the maximum number of neighbors of a node
increases, for all values of $q$ (notice that we have considered
that $q=0.3$ at most, namely a packet loss probability of 30\%).
This is rather intuitive, since as the number of neighbors
increases packet losses have less impact on the estimation and
thus better performance are achieved. Notice also that the value
of the function~\eqref{eq:coeff1_unif} for $q=0$ is
$1/|\mathcal{N}_{i}|$. Thus in presence of non-identical packet
loss probabilities the degradation in performance is not remarked.
In particular even when the first factor
of~\eqref{eq:expected_value_var-slightly-bounded} is very close to
1, if the number of neighbors is greater than 2, with a packet
loss of $q=0.3$ we have that the product of the two coefficient
does not exceed 0.65 and it is only a 30\% higher than the case
when no packet losses are present.
\begin{corollary} \label{cor:benchest}
Consider as benchmark the estimator computing the estimates by the
instantaneous average of the available measurements, namely the
estimator for which the weights are chosen to be $k_{ij}= 0$ and
$h_{ij} = 1/|\mathcal{N}_{\varphib_i}|$, for all $i = 1,\dots,N$,
and $j\in\mathcal{N}_{\varphib_i}$. Then, $\lim_{t\rightarrow
+\infty} \E_v e_i(t) = 0$ and the variance is
\begin{align}
    \E_\phi \E_v e_i^2 = \E_{\phib}
        \frac{\sigma^2}{\phib_i^T\phib_i}=\sigma^2\sum_{k = 0}^{|\mathcal{N}_{i}|-1} \frac{\chi(k)}{k+1}  \,.
    \label{eq:average_performance}
\end{align}
\end{corollary}

From this corollary we see that the difference in the expected
performance between the proposed estimator, given
by~\eqref{eq:expected_value_var-slightly-bounded}, and the
unbiased estimator that does an arithmetic average, given
by~\eqref{eq:average_performance}, is on the first coefficient
of~\eqref{eq:expected_value_var-slightly-bounded}. Clearly, the
proposed estimator outperforms the latter as the factor in~\eqref{eq:expected_value_var-slightly-bounded} is always
less than one, as shown in Figure~\ref{fig:e_phi_e_v_q}.

However, the bound~\eqref{eq:expected_value_var-slightly-bounded}
has been derived by
Proposition~\ref{eq:expected_value_var-slightly-bounded}, where
the cardinality of the set $\Theta_{\varphib_i}$ is bounded by
$N$. Obviously, this is in general a very conservative bound. The
set $\Theta_{\varphib_i}$ depends on the network topology, and no
tight bound can be derived unless some assumptions are given on
the network topology itself. We will show next that when we assume
information on the network topology, we are able to bound
$|\Theta_{\varphib_i}|$ more accurately. This further underlines
the improvement of the proposed estimator provides with respect to
the benchmark estimator of Corollary~\ref{cor:benchest}.

\begin{example}
Consider a simple line-graph. Let~$i$ be a node at the extreme of
the line-graph, then we have that $|\Theta_{\varphib_i}| \leq 2$.
Let~$j$ be a node of the line-graph different from the extremes,
then we have that $|\Theta_{\varphib_j}| \leq 3$. Thus
$$
    \sqrt{|\Theta_{\varphib_\ell}|^2+4} - |\Theta_{\varphib_\ell}| \geq
\left\{
  \begin{array}{ll}
    \sqrt{2^2+4}-2 \approx 0.828, & \hbox{if $\ell =i$;} \\
    \sqrt{3^2+4}-3 \approx 0.606, & \hbox{if $\ell =j$.}
  \end{array}
\right.
$$
By assuming that $\gamma_{\max} \in [0.5,1)$, we see that the
coefficient in~\eqref{eq:expected_value_var-slightly-bounded} is
at most $0.76$ for the border nodes and $0.708$ for those in the
middle, regardless the packet losses. Thus we have a significant
improvement with respect to the estimator that takes the average
of the measurements.


\end{example}

\begin{example}
Consider the family of finite Cayley graphs defined on a finite
additive Abelian group $(G,+)$, with
$G=\{0,1,\dots,|\mathcal{V}|-1\} = N$, namely the elements of the
group can be regarded as the labels of the nodes. The operator~$+$
is considered as addition modulo~$N$. Let us consider $S \subseteq
G$, such that $0\in S$ and $S$ is closed under the inverse, namely
if $a \in S$, then $-a \in S$. Two nodes, $i$~and~$j$, communicate
if and only if $(i-j) \mod N \in S$. Thus if $G =
\{0,1,2,\dots,N\}$ and $S = \{0,\pm 1,\pm 3, \pm 4\}$, then we
have a graph in which node~$i$ has as neighboring nodes those with
label $i\pm 1 \mod N$, $i \pm 3 \mod N$ and $i \pm 4 \mod N$. In
Figure~\ref{fig:example2a} a Cayley graph is shown, where $G =
\Z_{15}$.

\begin{figure}
    \centering
    \subfloat[]{\label{fig:example2a}\includegraphics[width=0.3\hsize]{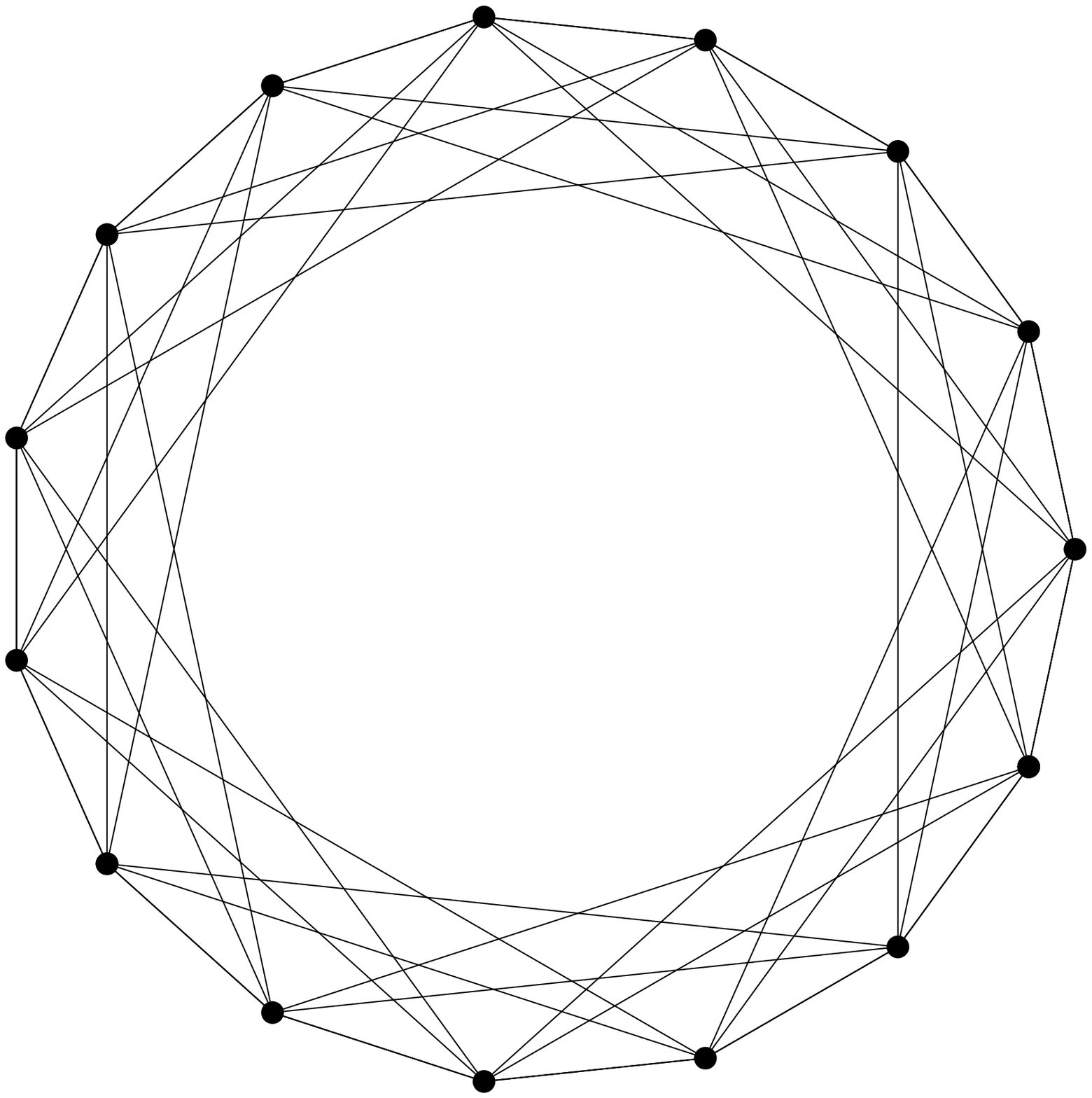}}\hspace*{2cm}
    \subfloat[]{\psfrag{x}[][]{$\gamma_{\max}$}
    \psfrag{b}[][]{$\nu$}
    \psfrag{y}[b][]{$\cfrac{(\sqrt{5}-1)\sqrt{\gamma_{\max}}+2(2\nu+1)}{2(\sqrt{5}-1)\sqrt{\gamma_{\max}}+2(2\nu+1)}$}
    \label{fig:example2b}\includegraphics[width=0.5\hsize]{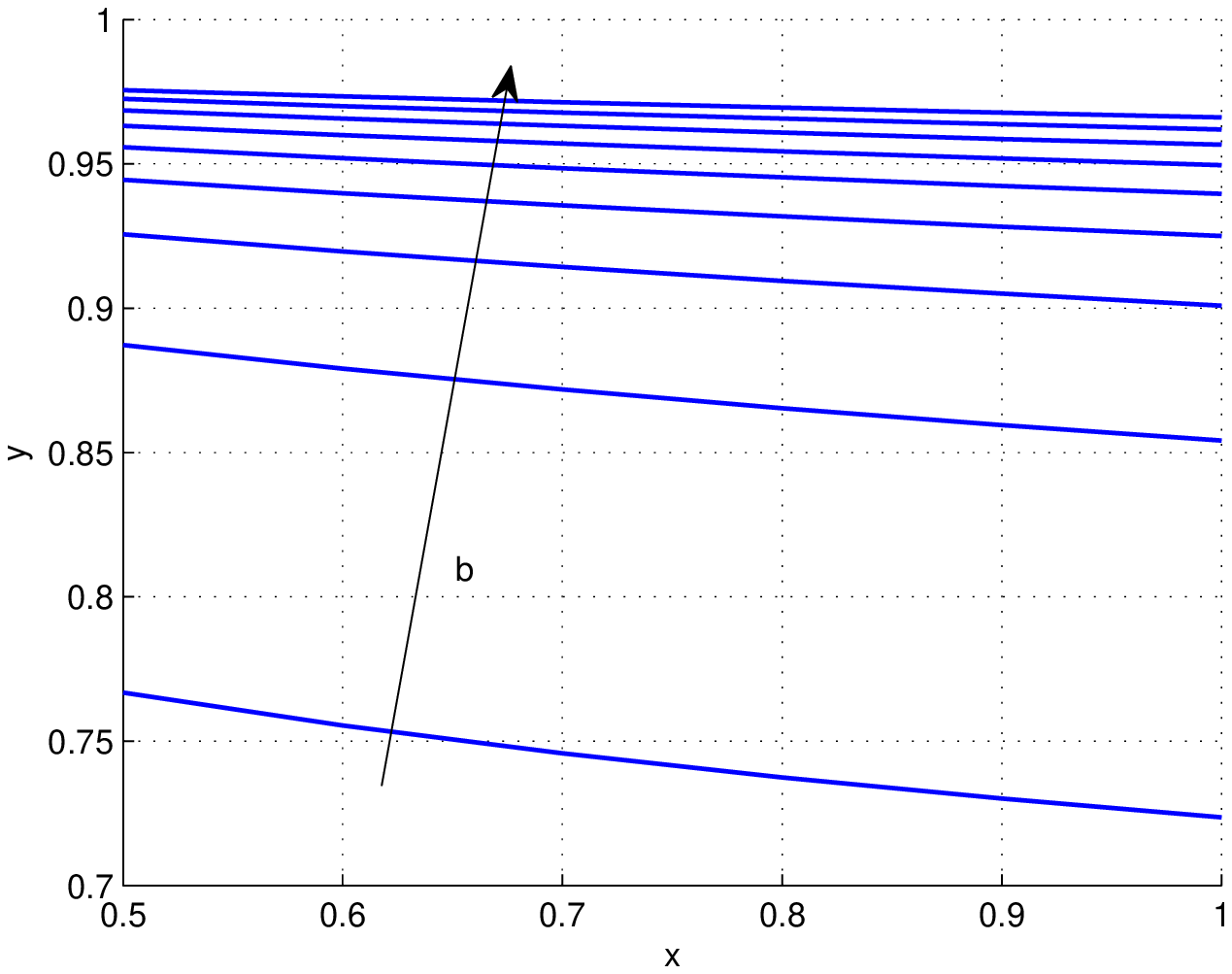}}
    \caption{On the left an example of Cayley graph $\mathcal{G} = (G,S)$
    defined on the group $G = \Z_{15}$ and $S=\{0,\pm 1, \pm 3, \pm 4\}$. On
    the right a plot that shows the first coefficient
    of~\eqref{eq:expected_value_var-slightly-bounded} as function of $\gamma_{\max}$ and $\nu$ when the network topology is described by a Cayley graph.
    Furthermore, $\gamma_{\max} \in [0.5,1)$ and $\nu \in [0,10]$. As it can be seen, the coefficient is always less then one, and,
    compared to Figure~\ref{fig:e_phi_e_v_q}, it depends only on~$\nu$, namely the degree of a node and not on the network size $N$.}
    \label{fig:example2}
\end{figure}

We have that each node communicates with $|S|-1= \nu$ nodes. In
other words, two distinct nodes have in common at most~$\nu$
nodes. This implies that $|\Theta_{\varphib_i}|\leq 2\nu+1$. Thus
$$
   \sqrt{|\Theta_{\varphib_i}|^2+4} - |\Theta_{\varphib_i}| \geq (\sqrt{5}-1) /(2\nu+1)\,.
$$
The first coefficient
of~\eqref{eq:expected_value_var-slightly-bounded} as function of
$\gamma_{\max}$ and $\nu$ is shown in Figure~\ref{fig:example2b}.
Notice that the function is similar in shape to that in
Figure~\ref{fig:e_phi_e_v_q}, however in this case the dependence
is on $\nu$, and not on the total number of nodes in the network
$N$. Albeit the network might have a total number of nodes many
orders of magnitude larger than $\nu$, the coefficient stays well
below the one in~\eqref{eq:expected_value_var-slightly-bounded},
which has been obtained when no information about the network is
known.

Since the coefficient
of~\eqref{eq:expected_value_var-slightly-bounded} is always much
less than one, the designed estimator outperforms significantly
the estimator that takes the average of the measurements.
\end{example}

%

\section{Simulations and Numerical Results} \label{sec:simulations}

Numerical simulations have been carried out to compare the
estimator proposed in this paper with some related estimators
available from the literature.

We consider the following five estimators:
\begin{description}
  \item[${E_1}$:] $\K(t)=\H(t)=(\I-\mathbf{L}(t))/2$ where $\mathbf{L}(t)$ is instantaneous Laplacian matrix
  associated to the graph $\mathcal{G}$. Clearly the graph changes when packets are dropped, so that arcs disappear from the graph.
  \item[${E_2}$:] $\K(t)=\0$ and $\H(t)=[h_{ij}(t)]$ with
      $h_{ij}=1/|\mathcal{N}_{\varphib_i}|$ if node $i$ and
      $j$ communicate, and $h_{ij}=0$ otherwise. Thus, the
      updated estimate is the  average of the measurements
      (this is the estimator of Corollary~\ref{cor:benchest}).
  \item[${E_3}$:] $\K(t)=[k_{ij}(t)]$, where $k_{ii}(t)=1/2|\mathcal{N}_{\varphib_i}|$,
$k_{ij}=1/|\mathcal{N}_{\varphib_i}|$ if node $i$ and $j$ communicate, $k_{ij}(t)=0$ otherwise, whereas $\H(t)=[h_{ij}(t)]$
  with $h_{ii}=1/2|\mathcal{N}_{\varphib_i}|$, and $h_{ij}=0$ elsewhere. This is the average of the old estimates and node's single measurement.
  \item[${E_4}$:] $\K(t)=\H(t)$ with
$k_{ij}(t)=1/2|\mathcal{N}_{\varphib_i}|$ if node $i$ and $j$
communicate, and $k_{ij}(t)=0$ otherwise. The updated estimate is the
average of the old estimates and all local new measurements.
   \item[${E_p}$:] The estimator proposed in this paper.
\end{description}
The estimators $E_1,\dots,E_4$ are based on various heuristics.
They are related to proposals in the literature, e.g., ${E_1}$
uses filter coefficients given by the Laplacian matrix,
cf.,~\cite{XiaoBoydLall}--\cite{Olfati05} and $E_2$ and $E_3$ are
considered here as benchmark. Observe that the weights based on
Laplacian do not ensure the minimization of the variance of the
estimation error.

\begin{figure}
    \centering
    \psfrag{d1}[][]{$d_1(t)$}
    \psfrag{d2}[][]{$d_2(t)$}
    \psfrag{d3}[][]{$d_3(t)$}
    \psfrag{d4}[][]{$d_4(t)$}
    \psfrag{d5}[][]{$d_5(t)$}
    \includegraphics[width=0.8\hsize]{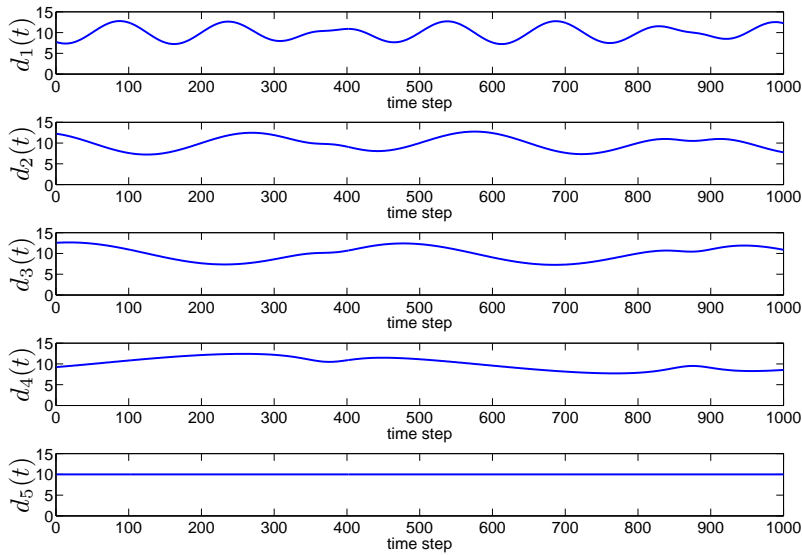}
    \caption{Test signals used in the simulations. Signal $d_2(t),\dots,d_5(t)$ are obtained from $d_1(t)$ by changing the frequency. The test
    signals are highly nonlinear and are generated so that the signal shows intervals in which the derivative (in absolute value) is
    small and intervals in which the derivative is higher.}
    \label{fig:dsig}
\end{figure}

Figure~\ref{fig:dsig} shows a set of test signals
$d_1(t),\dots,d_5(t)$ that has been used to assess the various
estimators. Note that the signals differ only in their frequency
content. The test signals are highly nonlinear and generated so
that the signal presents intervals in which is very slowly varying
(low absolute value of the derivative) and intervals in which the
derivative is higher.
The choice of the parameter $\gamma_{\max}$ is based on the
maximum cumulative bias, defined in
Equation~\eqref{eq:error_bound}: Let $P_b$ denote the desired
power of the cumulative biases of the estimates. Since there are
$N$ nodes, we consider the average power of the bias of each node
as $\Upsilon = P_b/N$. Assume that we want the estimator to
guarantee that the power of the right-hand side of
Equation~\eqref{eq:error_bound} is equal to $\Upsilon$. This is
equivalent to $$
  \gamma_{\max} = \frac{\sqrt{\Upsilon}}{\sqrt{\Upsilon}+\Delta}\,.
$$
In the simulations we have chosen $\Upsilon = 0\:$dB, which is a rather low value, and the noise variance is $\sigma^2 = 1.5$, which is quite a large value if compared to the amplitude of the signal to track. We also assume that he value of $\Delta$ is not known precisely but with an upper bound of about $5$\% from its real value. Therefore, these choices allow us to test the proposed estimator in the worst conditions.

\begin{figure}
  \centering
  \psfrag{X}[t][]{Packet loss probability $q$}
  \psfrag{Y}[b][]{MSE}
  \subfloat[]{\psfrag{T}[][]{MSE for estimation of $d_1(t)$}\includegraphics[width=0.45\hsize]{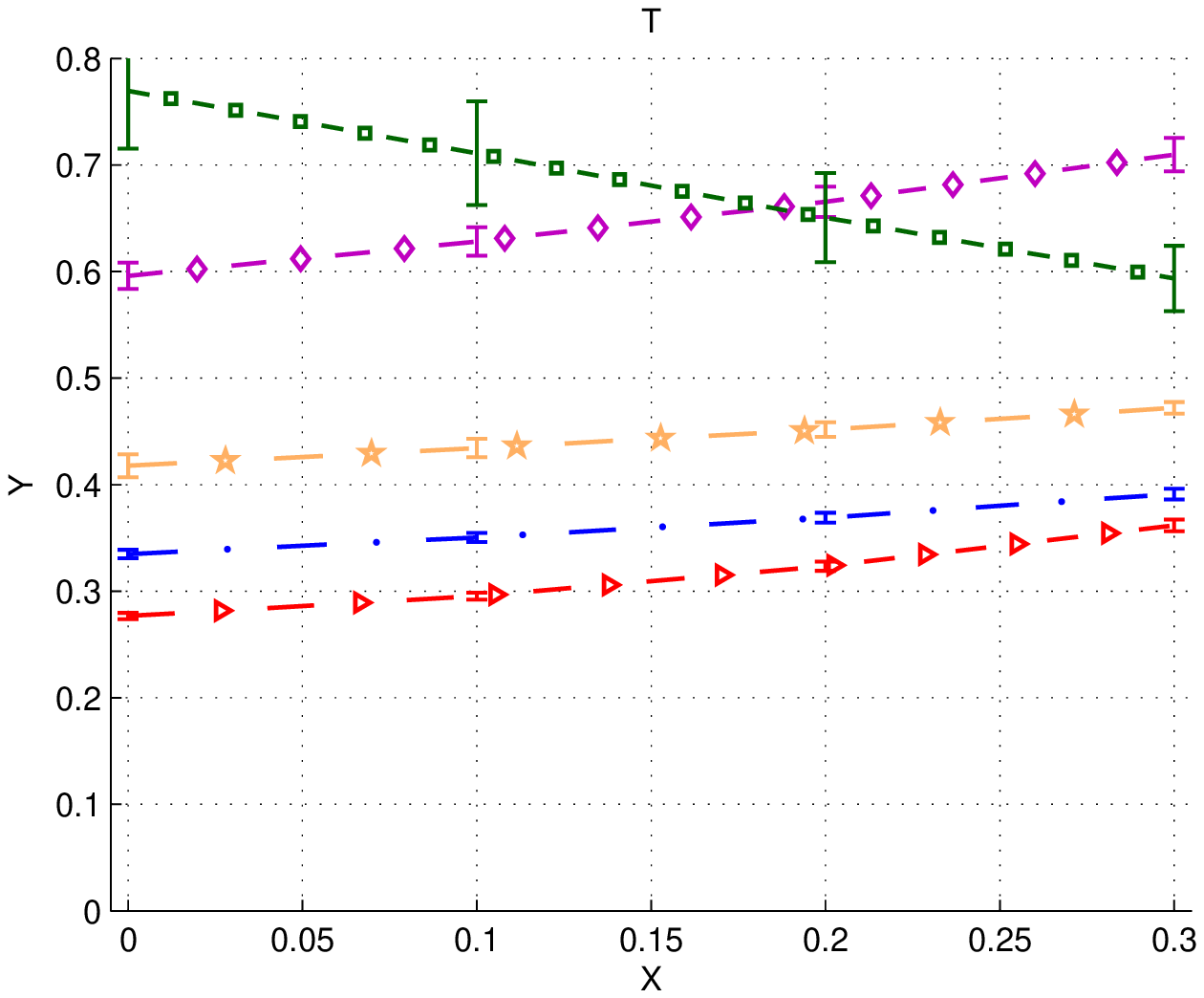}}
  \subfloat[]{\psfrag{T}[][]{MSE for estimation of $d_2(t)$}\includegraphics[width=0.45\hsize]{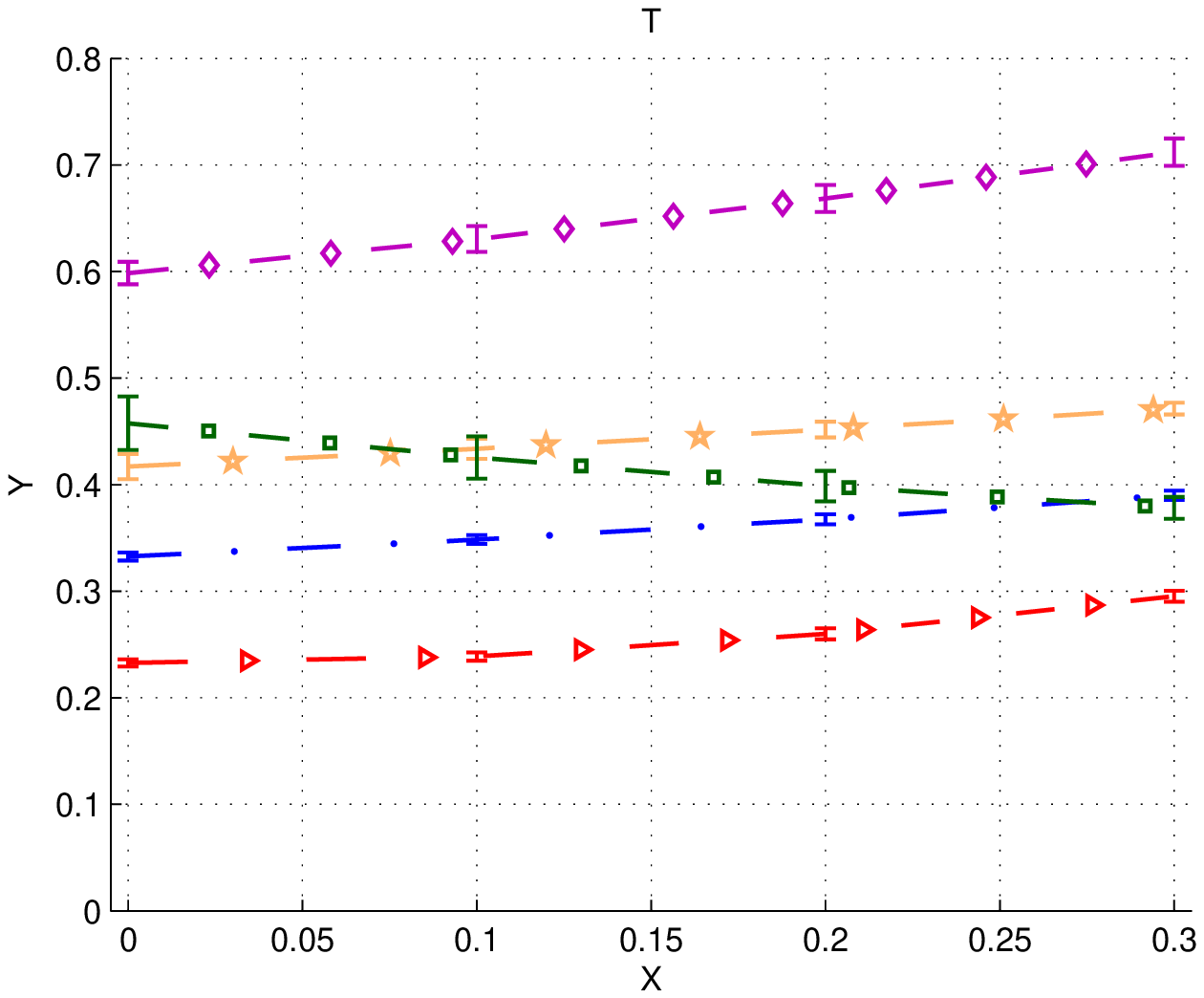}}\\
  \subfloat[]{ \psfrag{T}[][]{MSE for estimation of $d_3(t)$}\includegraphics[width=0.45\hsize]{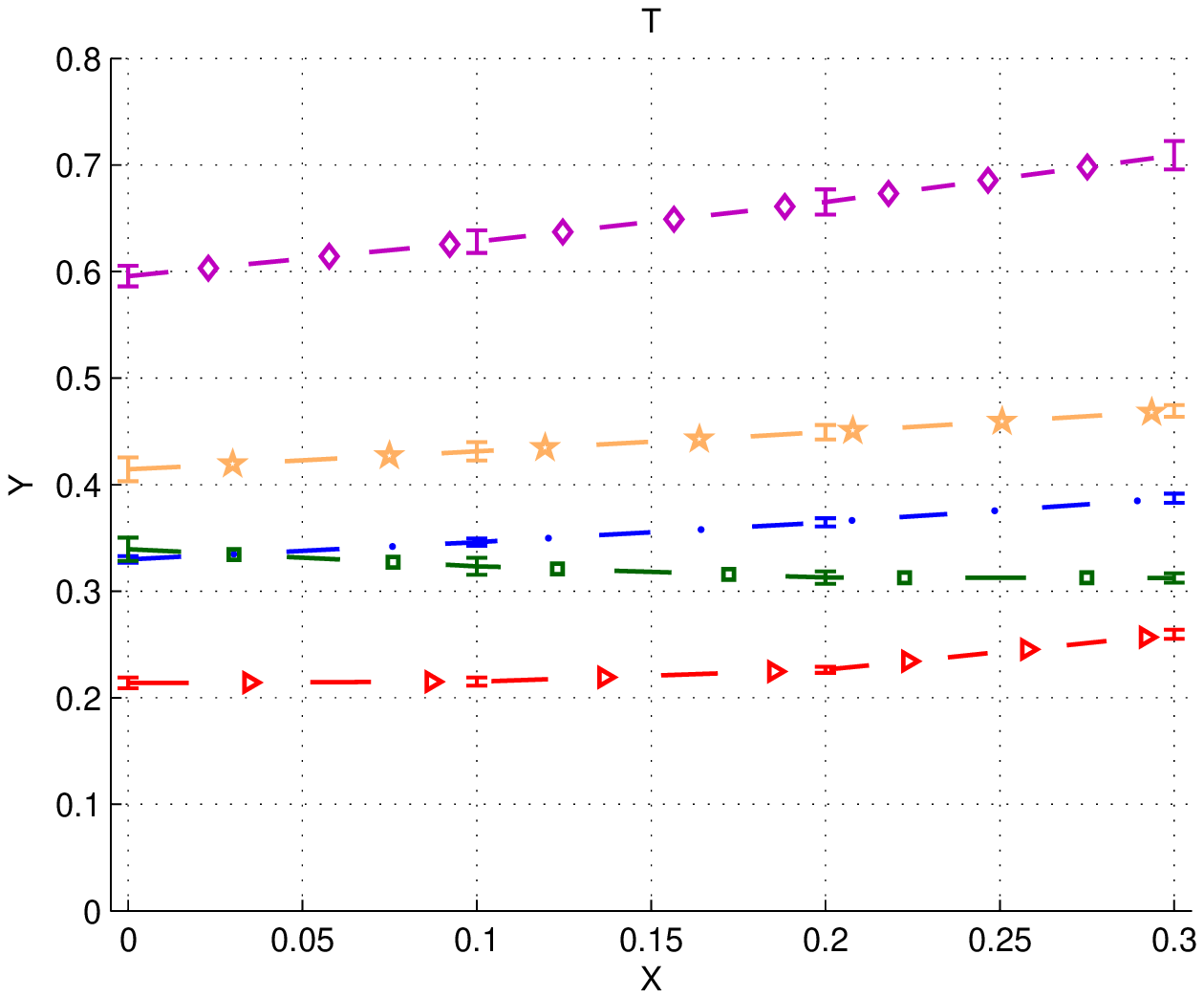}}
  \subfloat[]{  \psfrag{T}[][]{MSE for estimation of $d_4(t)$}\includegraphics[width=0.45\hsize]{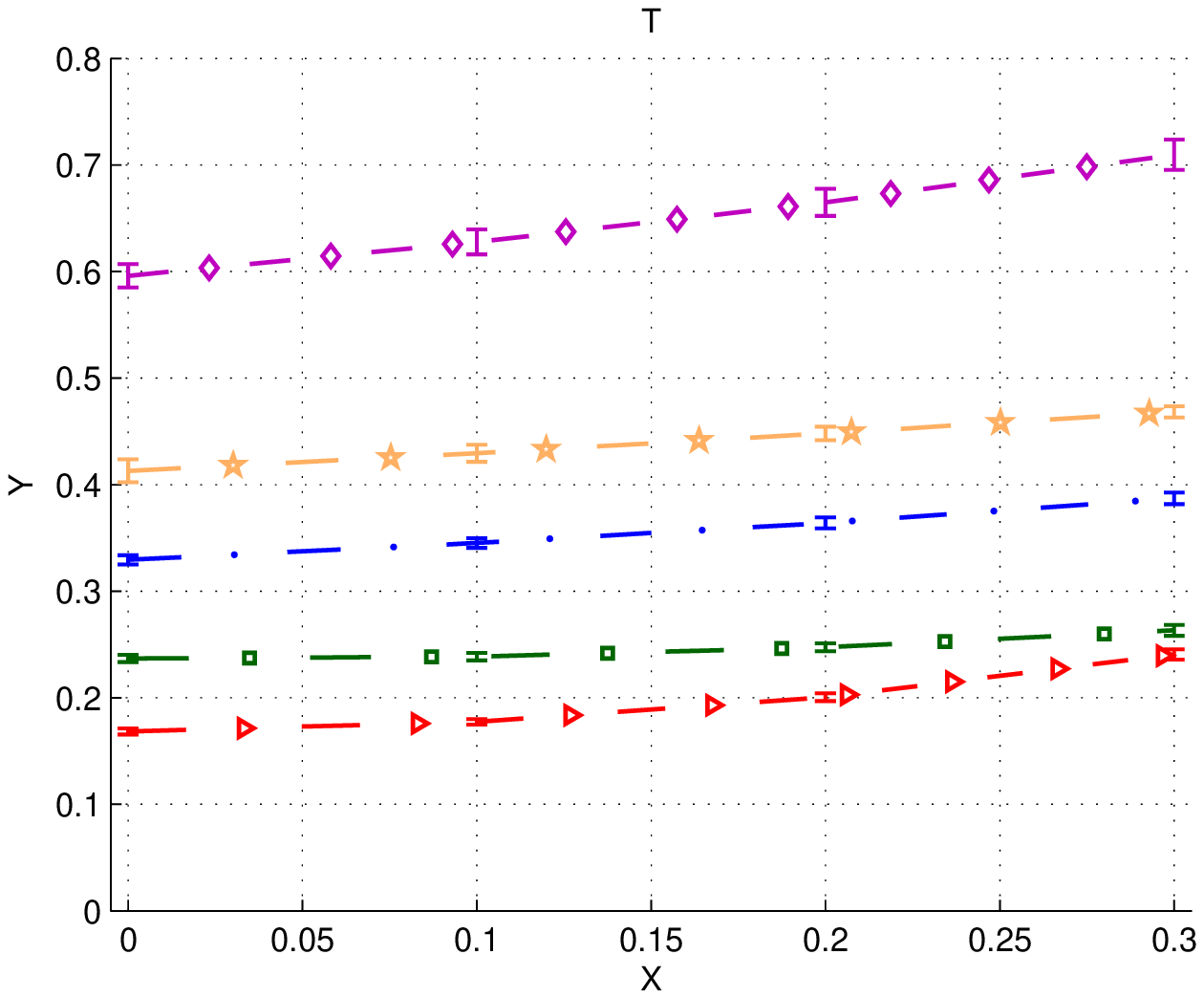}}\\
  \subfloat[]{\psfrag{T}[][]{MSE for estimation of $d_5(t)$}\includegraphics[width=0.45\hsize]{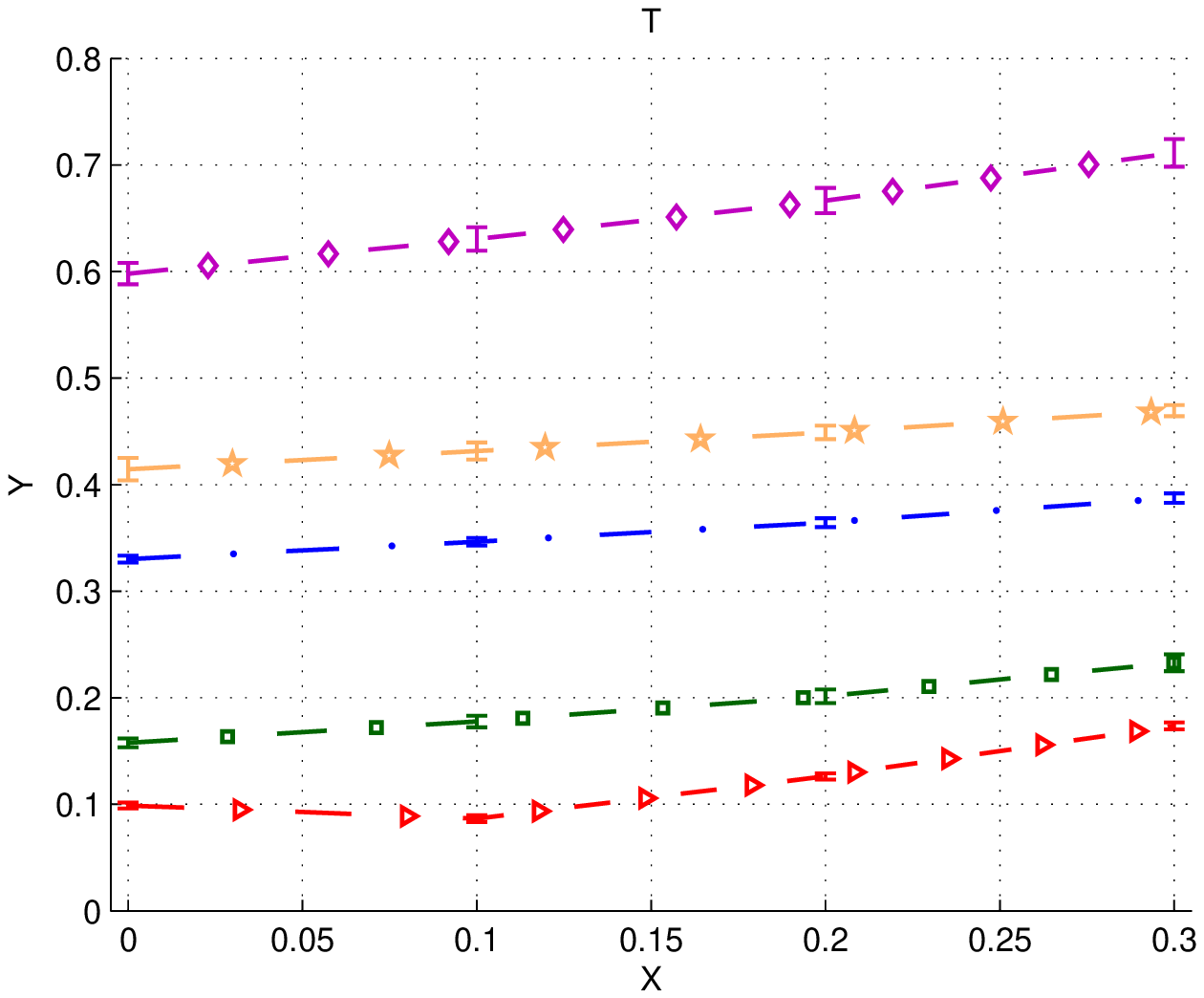}}
  \caption{Mean Square Error (MSE) performance comparison among estimators for various packet loss probabilities $q$. Each plot is associated to one of the five test signals $d_1(t),\dots,d_5(t)$, see Figure~\ref{fig:dsig}. The marker $(\star)$ refers to $E_1$, $(\lozenge)$ refers to $E_2$, $(\square)$ refers to $E_3$, $(\bullet)$ refers to $E_4$ and $({\Large\triangleright})$ refers to the proposed estimator $E_p$. The vertical bars represent the variance of the MSE computed for the 30 simulations. Notice that the proposed estimator $E_p$ has a very low variance, thus showing its high robustness to packet loss and measurement noise.}
  \label{fig:MSE_comparison}
\end{figure}

\begin{figure}
    \centering
    \psfrag{M}[][]{Measurements}
    \psfrag{E1}[][]{$E_1$}
    \psfrag{E2}[][]{$E_2$}
    \psfrag{E3}[][]{$E_3$}
    \psfrag{E4}[][]{$E_4$}
    \psfrag{Ep}[][]{$E_p$}
    \includegraphics[width=0.8\hsize]{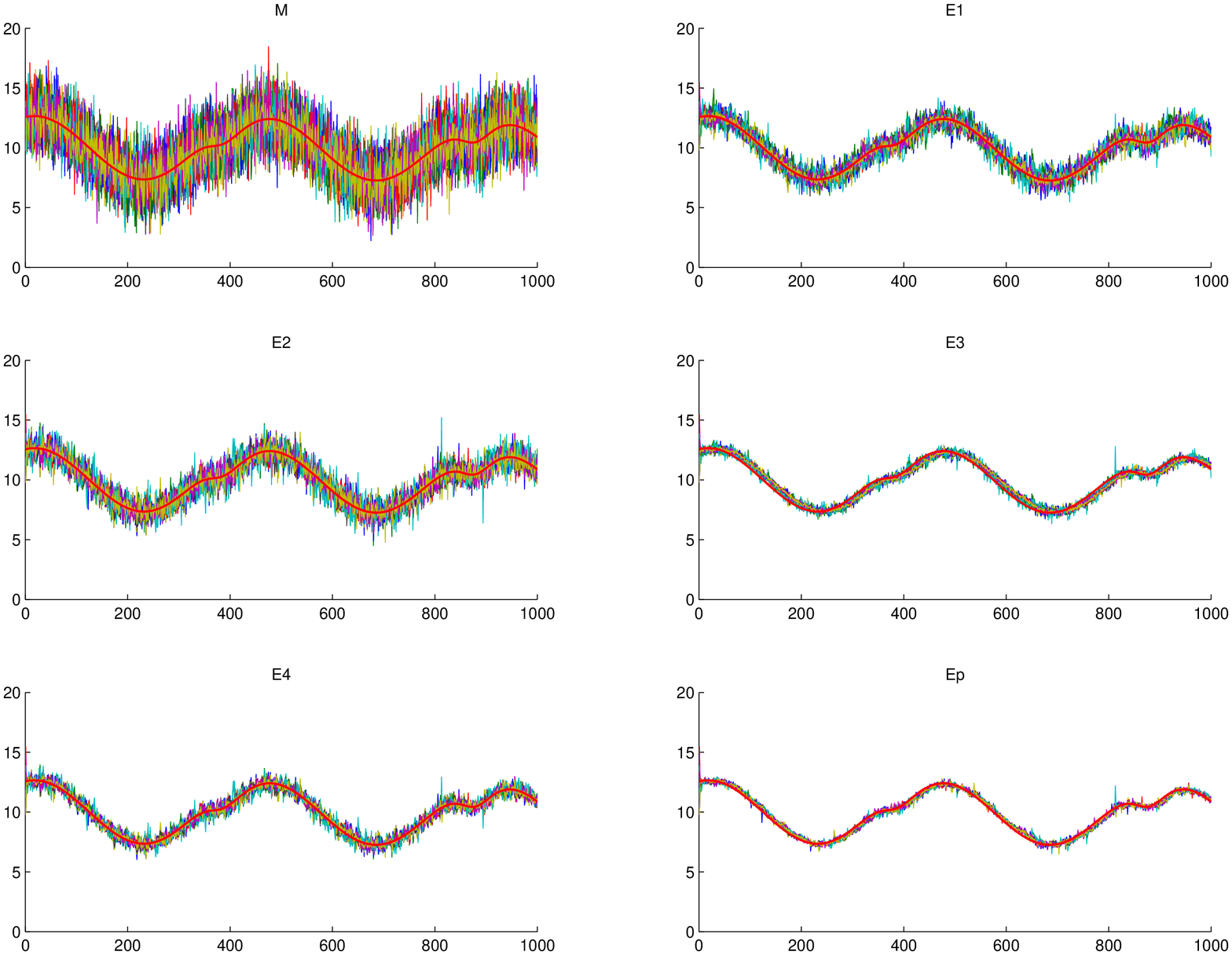}
    \caption{Realizations of the signal to be tracked, $d_2$, as it is measured by all the $N=20$ nodes, with a packet loss probability $q=10\%\pm5\%$. Notice that the proposed estimator $E_p$, visibly outperforms all the other estimators in term of variance. Notice also that the proposed estimator presents a small bias when the signal changes more rapidly.}
    \label{fig:curves}
\end{figure}

\begin{figure}
    \centering
    \psfrag{Y}[][]{$\gamma(K(t))$}
    \psfrag{svdk1}[][]{$\gamma(K(t))$ for $q=0.0$}
    \psfrag{svdk2}[][]{$\gamma(K(t))$ for $q=0.1$}
    \psfrag{svdk3}[][]{$\gamma(K(t))$ for $q=0.2$}
    \psfrag{svdk4}[][]{$\gamma(K(t))$ for $q=0.3$}
    \psfrag{time step}[][]{Time}
    \psfrag{a}[r][]{\small{From $d_5$ to $d_1$}}
    \includegraphics[width=0.95\hsize]{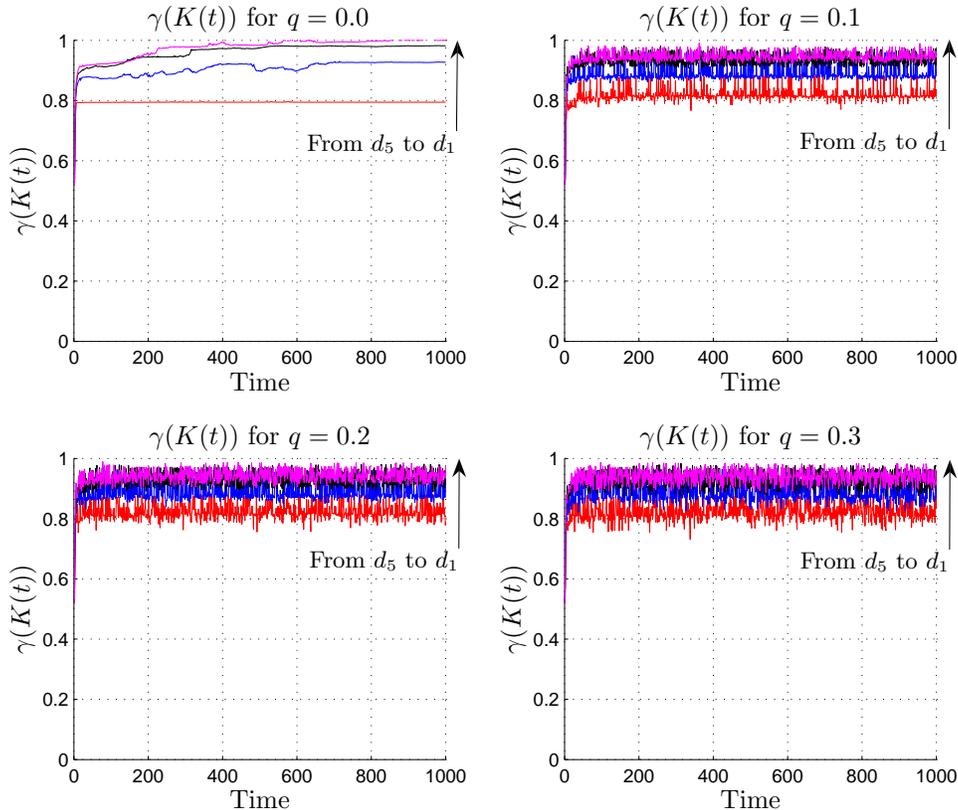}
    \caption{Largest value of the singular value of the matrix $\K(t)$ in the 30 Monte Carlo simulation generated. The rows of $\K(t)$ are computed individually by each node using~\eqref{eq:gainsk}. The curves correspond to the five test signals $d_1(t),\dots,d_5(t)$, and they are ordered from the slowest signal ($d_1(t)$) to the fastest one ($d_5(t)$).}
    \label{fig:svdk}
\end{figure}

We have considered 30 geometric random graphs with $N=20$ nodes
uniformly distributed in a square of side length equal to $N/2$.
Two nodes are connected if their Euclidean distance is less than
$1.7\sqrt{N}$. The average neighborhood size of all the considered
networks is $6.6$ nodes with a maximum and minimum neighborhood
size of $11$ and $3$, respectively. The estimation of the test
signals $d_1(t),\dots,d_5(t)$ has been performed under four
different packet loss probabilities. More precisely we have
consider the case in which $q_{ij}=0\%$, $q_{ij}=10\%\pm 5\%$,
$q_{ij}=20\%\pm 5\%$ and $q_{ij}=30\%\pm 5\%$. We also considered
different measurement noise realizations.
\def\MSE{\text{MSE}}
We take the mean square error of the estimates of each node  as
performance measure. Each estimator has an initial transition
phase, during which the mean square error may not be significant.
Hence, it has been computed after $70$ steps. Afterwards, the mean
square error has been averaged over all nodes of the network. The
average is denoted by $\MSE$. We define the improvement factor of
our estimator compared to the estimators $E_1,\ldots,E_4$ as $$
    \chi_i = \frac{\MSE({E_i}) -  \MSE({E_p})}{\MSE({E_i})}\,, \qquad i=1,\ldots,4\,.
$$
Figure~\ref{fig:MSE_comparison} shows the $\MSE({E_i})$ for all
the five different estimators as function of the packet loss
probability. In the simulation we assume the nodes know the
threshold value before the estimation process starts. This is
typically achieved after 10 time steps and thus the network experiences a short
initialization phase. Recall also that in all the simulations
related to the proposed estimator $E_p$, the error covariance
matrix is estimated at each node and reset as described in the end
of Section~\ref{sec:distrib_min_var_estim}.
Notice that $E_p$ outperforms all other estimators for any
considered packet loss probability. The two estimators $E_3$ and
$E_4$ have performance closer to $E_p$. In particular $E_3$
performs better when the signal is slow, whereas $E_4$ has better
performance when the signal is faster. This is motivated by the
fact that the estimator $E_3$ uses only one measurement and it is
affected by a bias that depends on the derivative of signal $d(t)$
(see Figure~\ref{fig:curves}). The $E_3$ performance improves as
the packet loss probability increases. Indeed, the single local
measurement is weighted more when packet losses occurs, i.e.,
previous estimates as lost, and thus the overall estimate becomes
less biased. The other two heuristic estimators, $E_2$ and $E_3$,
offer poor performance clearly in all the situations we have
considered.
Figure~\ref{fig:svdk} shows that the local computation of the weights $\k_i(t)$ with~\eqref{eq:gainsk} yields a stable $\K(t)$. In particular, higher values of $\gamma(\K(t))$ are experienced when the signal is slower, because $\Delta$ is small. Viceversa, lower values of $\gamma(\K(t))$ are obtained when $\Delta$ is large. This is explained by considering that when the signal is slow, then it is better to weight more previous estimates (which means larger $\K(t)$ and hence larger $\gamma(\K(t))$) to achieve small variances of the estimation error. By the same argument, when the signal is fast, then it is better to weight less previous estimates.

\section{Conclusions and Future Work}
\label{sec:conclusions} In this paper, we presented a
decentralized cooperative estimation algorithm for tracking a
time-varying signal using a wireless sensor network with lossy
communication. A mathematical framework was proposed to design a
filter, which run locally in each node of the network. Performance
analysis was carried out for the distributed estimation algorithm
in time varying communication networks with packet loss
probabilities both non-identical and identical among the links.
losses. We investigated how the estimation quality depends on
packet loss probability, network size and average number of
neighboring nodes. The theoretical analysis showed that the filter
is stable, and the variance of the estimation error is bounded
even in the presence of large packet loss probabilities. Numerical
results illustrated the validity of our approach, which
outperforms other estimators available from the literature.

Future studies will be devoted to the extension of our design methodology to the case when models of the signal to track are available. Lossy communication links with memory will also be included. Furthermore, we plan to implement our distributed filter on real wireless sensor networks, thus experimentally checking the validity of our theoretical predictions.

\section{Acknowledgments}
We would like to thanks the anonymous reviewers for the very useful comments that allowed us to improve and strengthen the paper.

\bibliographystyle{IEEEtran}
\bibliography{ref}



\end{document}